\pdfoutput=1
\documentclass[aps,prb,superscriptaddress,amsfonts,amsmath,amssymb,showpacs,floatfix,reprint,longbibliography]{revtex4-2}
\usepackage{multirow}
\usepackage{url}
\usepackage{bm}
\usepackage{graphicx}
\usepackage{amsmath}
\usepackage{amstext}
\usepackage{amssymb}
\usepackage{amsfonts}
\usepackage{amsbsy}
\usepackage{verbatim}
\usepackage{color}
\usepackage[colorlinks=true, urlcolor=blue, linkcolor=blue, citecolor=blue]{hyperref}
\usepackage{tikz}
\usetikzlibrary{arrows.meta}
\usepackage{gensymb}
\usepackage{textcomp}
\usepackage{enumitem}
\usepackage[version=3]{mhchem}
\usepackage{afterpage}
\usepackage{pbox}
\usepackage{makecell}
\usepackage{dsfont}
\usepackage{siunitx}
\usepackage{fancyhdr}
\usepackage{stackengine,wasysym,scalerel}
\usepackage{latexsym}
\usepackage{cancel}
\usepackage{array, bigdelim, booktabs}
\usepackage[misc]{ifsym}

\allowdisplaybreaks[1]

\begin{document}
\title{Projective symmetry groups beyond spin-$1/2$: Classifying integer-spin quantum spin liquids}

\author{Atanu Maity}
\affiliation{Institut f\"ur Theoretische Physik und Astrophysik and W\"urzburg-Dresden Cluster of Excellence ctd.qmat, Julius-Maximilians-Universit\"at W\"urzburg, Am Hubland, Campus S\"ud, W\"urzburg 97074, Germany}
\affiliation{Department of Physics, Indian Institute of Technology Madras, Chennai 600036, India}
\author{Chunxiao Liu}
\affiliation{Universit\'e Paris-Saclay, CNRS, Laboratoire de Physique des Solides, 91405 Orsay, France}
\author{Rhine Samajdar}
\affiliation{Department of Physics, Princeton University, Princeton, NJ 08544, USA}
\affiliation{Department of Electrical and Computer Engineering, Princeton University, Princeton, NJ 08544, USA}
\author{Yasir Iqbal}
\email{yiqbal@physics.iitm.ac.in}
\affiliation{Department of Physics, Indian Institute of Technology Madras, Chennai 600036, India}

\begin{abstract}
Wen's projective symmetry group (PSG), introduced a quarter of a century ago, reduced the classification of symmetric quantum spin liquids to an algebraic problem and has guided the study of spin-$1/2$ models and materials ever since. Many current candidate materials and quantum simulators, however, have spin-$1$ or larger moments, and no comparable classification exists for them. We formulate the PSG for fermionic partons of arbitrary spin $S$. Integer and half-odd-integer spins already differ in their gauge structure. Half-odd-integer spins give the familiar $\mathrm{SU}(2)$, $\mathrm{U}(1)$, and $\mathbb{Z}_2$ invariant gauge groups, while integer spins give a different hierarchy that includes $\mathrm{U}(1)\rtimes\mathbb{Z}_2$ and $\mathbb{Z}_2\times\mathbb{Z}_2$; the latter has no spin-$1/2$ counterpart. Three lattice-independent results follow for integer spin. Singlet pairing of the partons is odd under site exchange. A lattice symmetry can exchange the particle and hole representations of the local constraint, which forces symmetry-related sites into opposite constraint sectors. And a time-reversal-symmetric $\mathrm{U}(1)$ state with the same constraint on every site can be gapped only if the number of sites per unit cell is a multiple of $2S+1$. We apply the formalism to the spin-$1$ diamond lattice, realized in frustrated A-site spinels, and find ten fully symmetric \textit{Ans\"atze} up to second neighbors. In the $J_1$--$J_2$ Heisenberg model at the mean-field level, the ground state depends on how the constraint is imposed. With the particle and hole representations, the lowest states are fully gapped $\mathrm{U}(1)$ states whose two sublattices sit in opposite constraint sectors, and one of them is gauge equivalent to a paired \textit{Ansatz}. The classification provides a starting point for integer-spin quantum spin liquids on other lattices and in other materials.
\end{abstract}

\date{\today}
\maketitle
\pagestyle{myheadings}

\section{Introduction}
\label{sec:intro}
Not every phase of matter is captured by Landau's paradigm of spontaneous symmetry breaking. Quantum spin liquids (QSLs)~\cite{Anderson-1973} are the clearest example. In these magnets quantum fluctuations prevent order down to zero temperature, the excitations carry fractions of the quantum numbers of the spins, and they interact through emergent gauge fields, often with topological order~\cite{Wen-1991,Savary-2016,DallaPiazza-2015}. No local order parameter tells one QSL from another, so describing them required new concepts. Wen provided them a quarter of a century ago~\cite{Wen-2002}. Phases with the same symmetries can differ in how those symmetries act on the fractionalized excitations; Wen called this distinction quantum order and characterized it by the projective symmetry group (PSG).

When a spin-$1/2$ moment is written in terms of Abrikosov fermions, a microscopic symmetry need only be realized up to a gauge transformation of the partons. The PSG records how each symmetry is realized in this sense. It sorts the symmetric mean-field \textit{Ans\"atze} into classes with emergent $\mathrm{SU}(2)$, $\mathrm{U}(1)$, and $\mathbb{Z}_2$ gauge structure and allows one to enumerate them~\cite{Wen-2002}. In the twenty-five years since, this classification has become the usual way to go from a frustrated spin-$1/2$ Hamiltonian to a short list of candidate spin liquids, and from there to variational wave functions, low-energy field theories, and predictions for experiment. It has been applied to the kagome~\cite{Ran-2007,Lu-2011}, triangular~\cite{Lu-2016,Iqbal-2016}, and Kitaev honeycomb~\cite{You-2012} lattices in two dimensions and to the pyrochlore~\cite{Liu-2021} and diamond~\cite{Chauhan-2023} lattices in three.

All of this work concerns spin $1/2$, but many of the materials now studied as QSL candidates have spin $1$ or larger. Several frustrated spin-$1$ and higher-spin magnets show strongly suppressed ordering, broad excitation continua, field-induced dynamical regimes, or behavior close to a spin liquid. Spin-$1$ examples include the triangular magnet NiGa$_2$S$_4$~\cite{Nakatsuji-2005,Liu-2010b}, the pyrochlore fluoride NaCaNi$_2$F$_7$~\cite{Plumb-2019}, the langbeinite K$_2$Ni$_2$(SO$_4$)$_3$~\cite{Zivkovic-2021,Gonzalez-2024}, and diamond-lattice A-site spinels such as NiRh$_2$O$_4$~\cite{Chamorro-2018,Buessen-2018}. In spin-$3/2$ and spin-$5/2$ magnets such as CoAl$_2$O$_4$~\cite{MacDougall-2011}, MnSc$_2$S$_4$~\cite{Gao-2017}, KSrFe$_2$(PO$_4$)$_3$~\cite{Boya-2022}, and Li$_9$Fe$_3$(P$_2$O$_7$)$_3$(PO$_4$)$_2$~\cite{Kermarrec-2021}, frustration likewise suppresses or reshapes conventional order.
Synthetic quantum systems offer another route. Programmable arrays of neutral atoms have been used to study frustrated quantum magnets, including a $\mathbb{Z}_2$ topological spin liquid~\cite{Semeghini-2021,maity2024gapped,maity2024fermionic}. The same platforms can also encode higher spins: three Rydberg levels of a single atom form an effective spin $1$, which leads to dipolar spin-$1$ models with lattice supersolids and simplex phases on frustrated lattices~\cite{Liu-2024} and to the spin-$1$ Haldane phase in one dimension~\cite{Mogerle-2025}. In ultracold alkaline-earth atoms, the nuclear and electronic degrees of freedom decouple, which yields $\mathrm{SU}(N)$-symmetric magnets that are predicted to host spin-liquid ground states~\cite{Gorshkov-2010,CazalillaRey-2014}. In these platforms the spin can be chosen by design, so the question of which spin liquids are available to integer spins arises in quantum simulation as well as in materials.

With so many higher-spin candidates in both settings, a PSG classification for spin $S$ is now needed.
It cannot be obtained by simply adding parton flavors. Liu, Zhou, and Ng~\cite{Liu-2010a,Liu-2010b} introduced fermionic representations of arbitrary spin $S$, in which a spin-$S$ moment is described by $2S+1$ fermion flavors subject to a local constraint. For $S\!>\!1/2$ the particle-hole equivalence of spin $1/2$ is lost, and the internal redundancy depends on whether $S$ is integer or half-odd-integer: half-odd-integer spins keep an $\mathrm{SU}(2)$-type gauge structure, whereas integer spins have a $\mathrm{U}(1)\rtimes\mathbb{Z}_2$ structure~\cite{Liu-2010a}. In fermionic mean-field theory this difference underlies the even-odd effects of one-dimensional antiferromagnets. In two dimensions it allows new spin-$1$ spin liquids on the triangular lattice, including gapless $f$-wave and chiral $p_x+ip_y$ states~\cite{Liu-2010a,Liu-2010b}.

A related even-odd distinction appears in a different higher-spin parton construction. In the spin-$S$ Kitaev honeycomb model, the conserved plaquette operators remain exact $\mathbb{Z}_2$ gauge fluxes for any $S$, even though the model is not integrable for $S$\,$>$\,$1/2$~\cite{Ma-2023}. The corresponding $\mathbb{Z}_2$ gauge charges are fermions for half-odd-integer spin and bosons for integer spin. Fermionic charges must remain deconfined, while bosonic charges can condense and give a trivial paramagnet. The Kitaev construction uses Majorana partons rather than the representation used here, but it points to the same conclusion. Integer and half-odd-integer spins can differ already in their emergent gauge structure and in the nature of their fractionalized excitations.

Even so, there is still no counterpart of Wen's classification for $S\!>\!1/2$, and so no systematic way to list the symmetric spin liquids available to a higher-spin magnet. We construct this classification here for fermionic spin-$S$ partons. The invariant gauge groups (IGGs) depend on the parity of $2S$. Half-odd-integer spins give the familiar hierarchy of $\mathrm{SU}(2)$, $\mathrm{U}(1)$, and $\mathbb{Z}_2$ IGGs, while integer spins give $\mathrm{U}(1)\rtimes\mathbb{Z}_2$, $\mathrm{U}(1)$, $\mathbb{Z}_2$, and $\mathbb{Z}_2\times\mathbb{Z}_2$. The $\mathbb{Z}_2\times\mathbb{Z}_2$ IGG has no counterpart in the spin-$1/2$ Abrikosov-fermion classification and follows from the internal gauge structure of integer spins. Integer-spin QSLs can thus realize symmetries projectively in ways that spin-$1/2$ QSLs cannot.

The construction also has physical consequences that hold on any lattice. First, the singlet pairing field of integer-spin partons is odd under exchange of the two sites, whereas that of spin-$1/2$ Abrikosov fermions is even (Sec.~\ref{sec:spinS_formalism}); integer-spin paired states use different pairing channels. Second, for $S\!>\!1/2$ the particle and hole representations of the local constraint are inequivalent, and the gauge transformation that accompanies a lattice symmetry can exchange them. A symmetric state is then forced to place symmetry-related sites in opposite constraint sectors, unless it is self-dual (Appendix~\ref{app:selfdual}). Third, a counting argument valid for any spin gives a filling constraint. An integer-spin, time-reversal-symmetric $\mathrm{U}(1)$ state with all sites in the same constraint sector can be gapped only if the number of sites per unit cell is a multiple of $2S+1$ (Sec.~\ref{sec:filling}). Otherwise it has a spinon Fermi surface, and a gapped symmetric spin liquid must be paired or gauge equivalent to a state with sites in opposite sectors. None of these statements requires a calculation of energies.

As a first application we take the spin-$1$ diamond lattice, for two reasons. One is materials. A-site spinels, with magnetic ions on the diamond lattice, form a large family of frustrated three-dimensional magnets. The nearest-neighbor diamond-lattice antiferromagnet is bipartite, but a sizable second-neighbor exchange frustrates the two interpenetrating FCC sublattices and produces a degenerate surface of spiral ground states~\cite{Bergman-2007}. This mechanism has been used to explain the suppressed ordering and spiral correlations of several A-site spinels with $S=3/2$, $2$, and $5/2$. The spin-$1$ compound NiRh$_2$O$_4$ has motivated studies of spin-liquid behavior in spin-$1$ diamond antiferromagnets~\cite{Chamorro-2018,Buessen-2018}; pseudofermion functional renormalization group calculations find that the $J_1$--$J_2$ diamond antiferromagnet does not order for $S\leq1$ in the spiral regime, which points to a quantum spiral spin liquid for spin $1$~\cite{Buessen-2018}.

The other is gauge theory. The pyrochlore lattice is the medial lattice of the diamond lattice: pyrochlore sites sit on the bonds that connect neighboring tetrahedron centers. In quantum spin ice and related pyrochlore Coulomb phases, the spins can be viewed as variables on the links of a diamond lattice. The compact $\mathrm{U}(1)$ gauge field lives on these links, and the electric charges, or spinons, live on the diamond-lattice sites~\cite{Hermele-2004,Savary-2012,Lee-2012,Benton-2012}. A PSG always refers to a particular fractional excitation, and spinon PSGs describe how symmetries act on the spin-carrying partons. Spinon PSGs on the diamond lattice thus bear on pyrochlore magnets as well.

Our main results are the following.
\begin{enumerate}[leftmargin=*,label=(\roman*)]
\item \textit{A PSG classification for any spin.} We extend Wen's construction to fermionic partons of arbitrary spin $S$ (Sec.~\ref{sec:psW_general}). The gauge group of the partons, the hierarchy of invariant gauge groups, and the procedure for solving the PSG conditions are independent of the lattice. Hopping-only \textit{Ans\"atze} have the same link structure for integer and half-odd-integer spin, so existing spin-$1/2$ classifications can be used as the starting point for higher-spin ones.
\item \textit{A new gauge-group hierarchy for integer spin.} Integer spins realize the invariant gauge groups $\mathrm{U}(1)\rtimes\mathbb{Z}_2$, $\mathrm{U}(1)$, $\mathbb{Z}_2\times\mathbb{Z}_2$, and $\mathbb{Z}_2$ (Fig.~\ref{fig:fig1}). The $\mathbb{Z}_2\times\mathbb{Z}_2$ spin liquids have no spin-$1/2$ counterpart.
\item \textit{Lattice-independent constraints on integer-spin liquids.} Odd singlet pairing, the exchange of constraint sectors by lattice symmetries, and the filling constraint on gapped $\mathrm{U}(1)$ states (Sec.~\ref{sec:filling}) restrict which integer-spin liquids are possible on any lattice.
\item \textit{The complete short-range classification on the spin-$1$ diamond lattice.} Up to second-neighbor bonds there are ten fully symmetric \textit{Ans\"atze}, two of them with a $\mathbb{Z}_2\times\mathbb{Z}_2$ IGG (Sec.~\ref{sec:diamond_classification}); third-neighbor bonds add four $\mathbb{Z}_2$ descendants.
\item \textit{Energetics and experimental fingerprints.} In the spin-$1$ $J_1$--$J_2$ Heisenberg model the implementation of the local constraint decides the ground state. With the particle and hole representations, the lowest states are fully gapped $\mathrm{U}(1)$ states with the two sublattices in opposite constraint sectors, one of them gauge equivalent to a paired \textit{Ansatz} (Sec.~\ref{sec:j1j2_results}). We give their static structure factors, the quantities measured by neutron scattering, and compare them with the classical spiral surface (Sec.~\ref{sec:ssf}).
\end{enumerate}

The paper is organized as follows. Section~\ref{sec:spinS_formalism} reviews the fermionic representation of a spin-$S$ moment and the corresponding mean-field theory. Section~\ref{sec:psW_general} develops the PSG construction for spin-$S$ partons and the IGG hierarchies for integer and half-odd-integer spin. Section~\ref{sec:diamond_classification} applies it to the spin-$1$ diamond lattice and lists the fully symmetric short-range \textit{Ans\"atze}. Section~\ref{sec:j1j2_results} studies the spin-$1$ $J_1$--$J_2$ Heisenberg model within these \textit{Ans\"atze}, for the particle, hole, and particle-hole-symmetric implementations of the constraint and for the particle-hole-averaged representation of Ref.~\cite{Liu-2010a}. Section~\ref{sec:filling} derives a filling constraint on gapped $\mathrm{U}(1)$ states for general $S$, and Sec.~\ref{sec:ssf} discusses the static structure factors of the lowest states and compares them with the classical Luttinger-Tisza analysis. Section~\ref{sec:summary} summarizes the results.
\section{Fermionic spin-$S$ representation and mean-field construction}
\label{sec:spinS_formalism}
We first recall the fermionic representation of a spin-$S$ moment introduced in Ref.~\cite{Liu-2010a}. Each site carries $2S+1$ fermion flavors $f_m$, $m=-S,-S+1,\ldots,S$, and
\begin{equation}
S^\alpha = f_m^\dagger \Sigma^\alpha_{mn} f_n ,
\end{equation}
where repeated flavor indices are summed and $\Sigma^\alpha$ ($\alpha=x,y,z$) are the spin-$S$ matrices. The local Hilbert space is enlarged from the $(2S+1)$-dimensional spin space to a fermionic Fock space of dimension $2^{2S+1}$. There are two ways to recover the physical space, which look alike but are, in general, different. The particle representation imposes single occupancy,
\begin{equation}
N_f=\sum_{m=-S}^{S} f_m^\dagger f_m =1 ,
\end{equation}
and the hole representation imposes
\begin{equation}
N_f=2S ,
\end{equation}
i.e., a single hole among the $2S+1$ flavors. The two constraints coincide only for $S=1/2$, so the particle-hole equivalence of the spin-$1/2$ Abrikosov-fermion construction is lost for $S>1/2$~\cite{Liu-2010a}.

As in the spin-$1/2$ construction of Ref.~\cite{Affleck-1988}, we combine the fermions into the two-column object
\begin{equation}
\psi=(F,\bar F),
\end{equation}
where
\begin{equation}
F=\left(f_S,f_{S-1},\ldots,f_{-S}\right)^\mathrm{T},
\end{equation}
and
\begin{equation}
\bar F=
\big(f^\dagger_{-S},-f^\dagger_{-S+1},\ldots,
(-1)^{2S}f^\dagger_S \big)^\mathrm{T} .
\end{equation}
The spin operator then reads
\begin{equation}
\label{eq:spin_operator_two_spinor}
S^\alpha
=
\frac{1}{2}
{\rm Tr}\!\left[
\psi^\dagger
\Sigma^\alpha
\psi
\right],
\end{equation}
where $\Sigma^\alpha$ are the $(2S+1)\times(2S+1)$ spin-$S$
generators of $\mathrm{SU}(2)$, and the trace is over the two-column
internal space. In this form the internal redundancy of the
representation is easy to see. The spin operator is unchanged
under a right multiplication
\begin{equation}
\psi_i \rightarrow \psi_i W_i ,
\end{equation}
which suggests a $\mathrm{U}(2)$ gauge freedom acting on the two
columns of $\psi_i$. The two columns $F_i$ and $\bar F_i$ are,
however, not independent spinors, and only transformations that
preserve the canonical anticommutation relations are allowed. The
resulting internal symmetry group is
\begin{equation}
{\rm ISG}
=
\begin{cases}
\mathrm{SU}(2), & S\in\mathbb Z+\frac{1}{2},\\
\mathrm{U}(1)\rtimes\mathbb{Z}_2, & S\in\mathbb Z .
\end{cases}
\end{equation}
The internal structure of the representation thus differs
between half-odd-integer and integer spins, and this difference
runs through the whole spin-$S$ PSG classification.

The mean-field theory is built from spin-rotation-invariant link variables. A physical spin rotation acts from the left,
\begin{equation}
\psi_i \rightarrow G \psi_i ,
\end{equation}
where $G$ is the $(2S+1)$-dimensional irreducible representation of an $\mathrm{SU}(2)$ spin rotation, so the bilinear
\begin{equation}
\hat U_{ij}=\psi_i^\dagger \psi_j
\end{equation}
is a spin singlet. It is a $2\times2$ matrix in the internal space of $\psi$ and serves as the link variable on the bond $ij$. In terms of $F$ and $\bar F$ it contains the singlet hopping and pairing fields
\begin{equation}
\hat\chi_{ij}=F_i^\dagger F_j,\qquad
\hat\eta_{ij}=\bar F_i^\dagger F_j ,
\end{equation}
and takes the form
\begin{equation}
\hat U_{ij} =
\begin{pmatrix}
\hat\chi_{ij} & (-1)^{2S+1}\hat\eta^\dagger_{ij}  \\
\hat\eta_{ij} & -\hat\chi^\dagger_{ij}
\end{pmatrix}.
\label{eq:link}
\end{equation}
Under exchange of the two sites,
\begin{equation}
\hat\chi_{ji}=\hat\chi^\dagger_{ij},\qquad
\hat\eta_{ji}=(-1)^{2S+1}\hat\eta_{ij}.
\end{equation}
so the singlet pairing field is odd under site interchange for integer spin and even for half-odd-integer spin.

The local constraints are also simple in this notation. For integer spin, the particle and hole representations read
\begin{align}
{\rm Tr}\!\left[\psi_i\tau^3\psi_i^\dagger\right]
&= \zeta(2S-1), \nonumber\\
\zeta &=
\begin{cases}
+1, & N_f=1,\\
-1, & N_f=2S ,
\end{cases}
\label{eq:int_constraint}
\end{align}
where the Pauli matrix $\tau^3$ acts in the internal space of $\psi$. The $\mathrm{U}(1)$ part of $\mathrm{U}(1)\rtimes\mathbb{Z}_2$ preserves the choice of $\zeta$, while the $\mathbb{Z}_2$ part exchanges the particle and hole representations. For a fixed representation, one Lagrange multiplier $\lambda_i$ per site is enough to impose the constraint.

Three groups should be kept apart. (i) $\mathrm{U}(1)\rtimes\mathbb{Z}_2$ is the internal symmetry of the spin operator, i.e., of the unconstrained representation. (ii) Once an embedding, $\zeta=+1$ or $\zeta=-1$, is chosen on every site, only the $\mathrm{U}(1)$ subgroup preserves the constraint; the $\mathbb{Z}_2$ coset maps the particle theory onto the hole theory. (iii) In the particle-hole-averaged (or projected) formulation of Ref.~\cite{Liu-2010a}, which treats both embeddings alike, the full group acts as a gauge symmetry. The PSG classification below takes the gauge transformations from the full group, which is appropriate for the link sector and for the particle-hole-symmetric saddle point. At fixed integer filling, an \textit{Ansatz} whose PSG contains $\mathbb{Z}_2$-coset elements is symmetric only if the constraint label is transformed along with it, i.e., if sites related by such an element carry opposite embeddings. Section~\ref{sec:j1j2_results} works out what this means for the phase diagram.

For half-odd-integer spin the internal symmetry is $\mathrm{SU}(2)$, and the constraint is a vector in the internal space,
\begin{equation}
{\rm Tr}\!\left[\psi_i\boldsymbol{\tau}\psi_i^\dagger\right]
=
\bigl(0,0,\zeta(2S-1)\bigr),
\label{eq:half_odd_integer_constraint}
\end{equation}
where $\boldsymbol{\tau}=(\tau^1,\tau^2,\tau^3)$ acts in the
two-column internal space of $\psi$, and $\zeta=+1$ ($-1$)
corresponds to the particle (hole) representation.
An internal $\mathrm{SU}(2)$ rotation changes the direction of this vector, so three Lagrange multipliers $\boldsymbol{\lambda}_i=(\lambda_i^1,\lambda_i^2,\lambda_i^3)$ are needed, as in the spin-$1/2$ Abrikosov-fermion formulation.

At the mean-field level we decouple the spin-rotation-invariant channel with the $c$-number link variable
\begin{equation}
u_{ij}=\langle \hat U_{ij}\rangle .
\end{equation}
The quadratic Hamiltonian is
\begin{align}
\hat H_{\rm MF}
=&\;
\frac{a(S)}{2}\sum_{ij}J_{ij}
\Bigg\{
{\rm Tr}\!\left(u_{ij}^\dagger u_{ij}\right)
\nonumber\\
&\qquad
-
{\rm Tr}\!\left[
u_{ij}^\dagger \psi_i^\dagger\psi_j
+{\rm H.c.}
\right]
\Bigg\}
+\hat H_c ,
\label{eq:quadratic_ham}
\end{align}
where $a(S)$ is a spin-dependent coefficient, the sum runs over ordered pairs $(i,j)$, so that each bond is counted twice, and the trace is over the internal indices. For $S=1$ the Heisenberg exchange can be written exactly as $J\,\mathbf S_i\cdot\mathbf S_j=-J:\!(\hat\chi^\dagger_{ij}\hat\chi_{ij}+\hat\eta^\dagger_{ij}\hat\eta_{ij})\!:$~\cite{Liu-2010a}. Equation~\eqref{eq:quadratic_ham} with $a(1)=1/2$ is then the complete mean-field decoupling of the Heisenberg model. For $S>1$ the exchange also contains triplet and higher-multipole channels~\cite{Liu-2010a}, and Eq.~\eqref{eq:quadratic_ham} is a spin-rotation-invariant trial Hamiltonian in the singlet channel only. For integer spin the constraint term is
\begin{equation}
\hat H_c
=
\sum_i \lambda_i
\left[
{\rm Tr}\!\left(\psi_i\tau^3\psi_i^\dagger\right)
-\zeta(2S-1)
\right],
\label{eq:constraint_integer}
\end{equation}
and for half-odd-integer spin
\begin{equation}
\hat H_c
=
\sum_i
\boldsymbol{\lambda}_i\cdot
\left[
{\rm Tr}\!\left(\psi_i\boldsymbol{\tau}\psi_i^\dagger\right)
-
\bigl(0,0,\zeta(2S-1)\bigr)
\right].
\label{eq:constraint_half_odd_integer}
\end{equation}
Here $\zeta=+1$ and $-1$ denote the particle and hole representations.

The redundancy of the representation acts as a local gauge transformation. Under
\begin{equation}
\psi_i\rightarrow \psi_i W_i,\qquad W_i\in{\rm ISG},
\end{equation}
the mean-field link and constraint fields transform as
\begin{equation}
u_{ij}\rightarrow W_i^\dagger u_{ij}W_j .
\end{equation}
For integer spin,
\begin{equation}
\lambda_i\tau^3 \rightarrow
W_i^\dagger \lambda_i\tau^3 W_i ,
\end{equation}
whereas for half-odd-integer spin,
\begin{equation}
\boldsymbol{\lambda}_i\cdot\boldsymbol{\tau}
\rightarrow
W_i^\dagger
\big(\boldsymbol{\lambda}_i\cdot\boldsymbol{\tau}\big)
W_i .
\end{equation}
Equation~\eqref{eq:quadratic_ham} is invariant under these transformations. This is the starting point of the PSG construction: a physical symmetry need not leave an \textit{Ansatz} invariant by itself, as long as it does so when combined with a gauge transformation.
\section{Projective symmetry group for spin-$S$ partons}
\label{sec:psW_general}
At the saddle point the operator-valued link field is replaced by its expectation value,
$
u_{ij}=\langle \hat U_{ij}\rangle
$, and
the onsite constraint is imposed only on average through a Lagrange multiplier field. We write the corresponding matrix in the internal space as
\begin{equation}
\Lambda_i =
\begin{cases}
\lambda_i\tau^3,
& S\in\mathbb{Z},\\
\boldsymbol{\lambda}_i\cdot\boldsymbol{\tau},
& S\in\mathbb{Z}+\frac12 ,
\end{cases}
\label{eq:Lambda_def}
\end{equation}
where $\boldsymbol{\tau}=(\tau^1,\tau^2,\tau^3)$ acts on the two columns of $\psi$. A mean-field \textit{Ansatz} is specified by the set
\begin{equation}
\mathcal A=\{u_{ij},\Lambda_i\}.
\end{equation}
\textit{Ans\"atze} related by a local ISG transformation describe the same physical spin state before projection. This local ISG redundancy is why symmetries can be realized projectively.

Let $\mathcal O$ be a symmetry of the lattice space group, with the convention that site $i$ is mapped to $\mathcal O(i)$.  Acting with $\mathcal O$ alone sends the \textit{Ansatz} to
\begin{equation}
\mathcal O:\{u_{ij},\Lambda_i\}
\longrightarrow
\{u_{\mathcal O(i)\mathcal O(j)},\Lambda_{\mathcal O(i)}\}.
\end{equation}
The \textit{Ansatz} is symmetric under $\mathcal O$ if the transformed \textit{Ansatz} is gauge-equivalent to the original one, i.e., if there is a site-dependent internal transformation $W_{\mathcal O}$ such that
\begin{equation}\label{eq:projective_condition}
\begin{aligned}
W_{\mathcal O}^\dagger[\mathcal O(i)]\,
u_{\mathcal O(i)\mathcal O(j)}\,
W_{\mathcal O}[\mathcal O(j)]
&= u_{ij},\\
W_{\mathcal O}^\dagger[\mathcal O(i)]\,
\Lambda_{\mathcal O(i)}\,
W_{\mathcal O}[\mathcal O(i)]
&= \Lambda_i .
\end{aligned}
\end{equation}
The pair $(\mathcal O,W_{\mathcal O})$ is then an element of the projective symmetry group. A physical symmetry thus need not act linearly on the partons; it may come with a gauge transformation in the internal space.

Time reversal is antiunitary and needs separate treatment. We write
\begin{equation}
\mathcal T=\mathcal U_{\mathcal T}\mathcal K,
\qquad
\mathcal U_{\mathcal T}=e^{-i\pi S^y},
\end{equation}
where $\mathcal K$ denotes complex conjugation.  The spin matrices obey
\begin{equation}
\mathcal T S_i^\alpha \mathcal T^{-1}=-S_i^\alpha .
\end{equation}
On the two-column fermion field, time reversal acts as
\begin{equation}
\psi_i\rightarrow \mathcal U_{\mathcal T}\mathcal K\psi_i .
\end{equation}
In a projective implementation this operation may be followed by an internal transformation,
\begin{equation}
\psi_i\rightarrow
\mathcal U_{\mathcal T}\mathcal K\psi_i\,W_{\mathcal T}(i).
\end{equation}
To obtain the condition on the link matrix, consider the bilinear
\begin{equation}
{\rm Tr}\!\left(\psi_i u_{ij}\psi_j^\dagger\right).
\end{equation}
Under projective time reversal it transforms as
\begin{equation}
\begin{split}
{\rm Tr}\!\left(\psi_i u_{ij}\psi_j^\dagger\right)
\rightarrow
{\rm Tr}\!\left[
\mathcal U_{\mathcal T}\psi_i
W_{\mathcal T}^\ast(i)u_{ij}^\ast
W_{\mathcal T}^{T}(j)
\psi_j^\dagger
\mathcal U_{\mathcal T}^\dagger
\right].
\end{split}
\end{equation}
The complex conjugation of the gauge matrix can be removed using
\begin{equation}
W^\ast=\tau^\nu W\tau^\nu ,
\qquad
\nu=
\begin{cases}
1, & S\in\mathbb{Z},\\
2, & S\in\mathbb{Z}+\frac12,
\end{cases}
\end{equation}
with $W\in \mathrm{U}(1)\rtimes\mathbb{Z}_2$ for integer spin and $W\in \mathrm{SU}(2)$ for half-odd-integer spin.  A spin-rotation-invariant link matrix can be written as
\begin{equation}\label{eq:mean_link}
u_{ij}
=
i\lambda_{ij}^0\tau^0
+\lambda_{ij}^3\tau^3
+i^p\lambda_{ij}^1\tau^1
+i^p\lambda_{ij}^2\tau^2 ,
\end{equation}
where $p=(2S+1)\bmod 2$ and all $\lambda_{ij}^\gamma$ are real.  The $\tau^0$ and $\tau^3$ components are singlet hopping amplitudes and the $\tau^1$ and $\tau^2$ components are singlet pairing amplitudes; in the canonical \textit{Ans\"atze} of the appendices, these are denoted $\chi^0_{ij}=\lambda_{ij}^0$, $\chi^3_{ij}=\lambda_{ij}^3$, $\Delta^1_{ij}=\lambda_{ij}^1$, and $\Delta^2_{ij}=\lambda_{ij}^2$.  With this convention,
\begin{equation}
\tau^\nu u_{ij}^\ast \tau^\nu=-u_{ij}.
\end{equation}
Using also $W^{T}=(W^\ast)^\dagger=\tau^\nu W^\dagger\tau^\nu$, the transformed bilinear becomes $-{\rm Tr}[\mathcal U_{\mathcal T}\psi_i\,(\tau^\nu W_{\mathcal T}(i))\,u_{ij}\,(\tau^\nu W_{\mathcal T}(j))^\dagger\psi_j^\dagger\mathcal U_{\mathcal T}^\dagger]$. We absorb the extra $\tau^\nu$ into the gauge part of time reversal by defining
\begin{equation}
\widetilde W_{\mathcal T}(i)=
\begin{cases}
\big[\tau^1 W_{\mathcal T}(i)\big]^\dagger=W^\dagger_{\mathcal T}(i)\tau^1,
& S\in\mathbb{Z},\\
\big[i\tau^2 W_{\mathcal T}(i)\big]^\dagger=-iW^\dagger_{\mathcal T}(i)\tau^2,
& S\in\mathbb{Z}+\frac12 .
\end{cases}
\end{equation}
With this gauge matrix the time-reversal condition on the \textit{Ansatz} reads
\begin{equation}\label{eq:projective_condition_time}
\begin{aligned}
\widetilde W_{\mathcal T}^\dagger(i)\,
u_{ij}\,
\widetilde W_{\mathcal T}(j)
&= -u_{ij},\\
\widetilde W_{\mathcal T}^\dagger(i)\,
\Lambda_i\,
\widetilde W_{\mathcal T}(i)
&= -\Lambda_i .
\end{aligned}
\end{equation}
The minus sign comes from the antiunitarity of time reversal together with our convention for the singlet link variable. Equation~\eqref{eq:projective_condition_time} is the time-reversal analog of the unitary PSG condition in Eq.~\eqref{eq:projective_condition}. In the appendices the gauge part of time reversal is written as $W_{\mathcal T}(r,\mu)$; it is the matrix $\widetilde W_{\mathcal T}$ of Eq.~\eqref{eq:projective_condition_time}, so that all PSG conditions involving $\mathcal T$ are imposed on $\widetilde W_{\mathcal T}$ with the minus sign of Eq.~\eqref{eq:projective_condition_time}.

The invariant gauge group (IGG) is the group of pure gauge transformations that leave the \textit{Ansatz} unchanged, i.e., the PSG elements whose physical part is the identity. An element $\mathcal W_i\in{\rm IGG}$ satisfies
\begin{equation}\label{eq:igW_definition}
\begin{aligned}
\mathcal W_i^\dagger u_{ij}\mathcal W_j
&= u_{ij},\\
\mathcal W_i^\dagger \Lambda_i\mathcal W_i
&= \Lambda_i .
\end{aligned}
\end{equation}
The IGG is the kernel of the projection of the PSG onto the physical symmetry group. Equivalently, algebraic relations among the physical symmetries need to hold only up to elements of the IGG. Since $\Lambda_i$ is part of the \textit{Ansatz} and $\tau^1\Lambda_i\tau^1=-\Lambda_i$, a nonzero onsite field removes the $\mathbb{Z}_2$ coset from the IGG of the link sector, $\mathrm{U}(1)\rtimes\mathbb{Z}_2\to\mathrm{U}(1)$ and $\mathbb{Z}_2\times\mathbb{Z}_2\to\mathbb{Z}_2$; the IGGs quoted in the classification below refer to the link sector.

\begin{figure}
\centering
\includegraphics[width=1.0\linewidth]{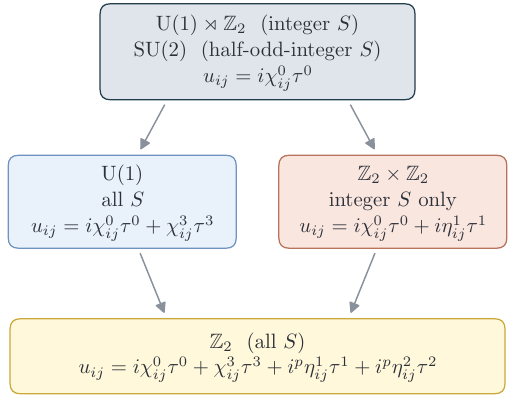}
\caption{Invariant gauge groups (IGGs) of the link sector and the corresponding canonical link matrices in the fermionic spin-$S$ construction. Arrows indicate which components of $u_{ij}$ must be switched on to reduce the IGG. For half-odd-integer spin the hierarchy is $\mathrm{SU}(2)\to\mathrm{U}(1)\to\mathbb{Z}_2$; for integer spin the top group is $O(2)\simeq\mathrm{U}(1)\rtimes\mathbb{Z}_2$ and the additional $\mathbb{Z}_2\times\mathbb{Z}_2$ level appears. The integer $p=(2S+1)\bmod2$ fixes the factors of $i$ in the pairing components. A nonzero onsite field $\Lambda_i=\lambda_i\tau^3$ further reduces $\mathrm{U}(1)\rtimes\mathbb{Z}_2\to\mathrm{U}(1)$ and $\mathbb{Z}_2\times\mathbb{Z}_2\to\mathbb{Z}_2$ (see text).}
\label{fig:fig1}
\end{figure}

The IGG is determined by which link components an \textit{Ansatz} contains. In a canonical gauge the IGG is a subgroup of the global internal symmetry group; the canonical forms are shown in Fig.~\ref{fig:fig1}. For half-odd-integer spin the possible IGGs are
\begin{equation}
\mathrm{SU}(2),\qquad \mathrm{U}(1),\qquad \mathbb{Z}_2 ,
\end{equation}
with representative global gauge matrices
\begin{equation}
\left\{e^{i\phi\,\hat{\mathbf n}\cdot\boldsymbol{\tau}}\right\},
\qquad
\left\{e^{i\phi\tau^3}\right\},
\qquad
\left\{\pm \tau^0\right\},
\end{equation}
respectively.  For integer spin, the corresponding possibilities are
\begin{equation}
\mathrm{U}(1)\rtimes\mathbb{Z}_2,\qquad
\mathrm{U}(1),\qquad
\mathbb{Z}_2,\qquad
\mathbb{Z}_2\times\mathbb{Z}_2 ,
\end{equation}
with canonical representatives
\begin{equation}
\begin{aligned}
&\left\{e^{i\phi\tau^3},\,e^{i\phi\tau^3}\tau^1\right\},
\qquad
\left\{e^{i\phi\tau^3}\right\},\\
&\left\{\pm\tau^0\right\},
\qquad
\left\{\pm\tau^0,\,\pm\tau^1\right\}.
\end{aligned}
\end{equation}
Here, $\phi$ is a real parameter, $\hat{\mathbf n}$ is a unit vector in the internal $\mathrm{SU}(2)$ space, and $\tau^0$ is the $2\times2$ identity matrix.  The first integer-spin group is the orthogonal group $O(2)\simeq\mathrm{U}(1)\rtimes\mathbb{Z}_2$: the element $\tau^1$ acts on the continuous subgroup by inversion, $\tau^1e^{i\phi\tau^3}\tau^1=e^{-i\phi\tau^3}$, so the product is semidirect rather than direct. The last group is the Klein four-group.

The exchange property of the hopping field has a simple consequence. Since
$
\chi_{ji}=\chi_{ij}^\dagger
$
for both integer and half-odd-integer spins, hopping-only \textit{Ans\"atze} have the same link structure in the two cases. Their IGGs can still differ, because the allowed gauge transformations differ.  For example, the canonical \textit{Ansatz} with only the $\lambda_{ij}^0$ component nonzero has IGG $\mathrm{U}(1)\rtimes\mathbb{Z}_2$ for integer spin, whereas the corresponding half-odd-integer \textit{Ansatz} has IGG $\mathrm{SU}(2)$, as shown in Fig.~\ref{fig:fig1}.  Integer and half-odd-integer spins thus differ both in the parity of the pairing field and in the group of allowed gauge transformations.

For an \textit{Ansatz} that is not in the canonical gauge, the IGG is most easily found from gauge-covariant flux operators.  Let $\mathcal C_j$ be a closed loop based at site $j$ and passing through the ordered sequence of sites
\begin{equation}
\mathcal C_j=(j,k,l,\ldots,m,j).
\end{equation}
The flux operator is
\begin{equation}
P_{\mathcal C_j}
=
P(j,k,l,\ldots,m,j)
=
u_{jk}u_{kl}\cdots u_{mj}.
\label{eq:flux_def}
\end{equation}
Under a local gauge transformation $u_{ij}\rightarrow W_i^\dagger u_{ij}W_j$, this object transforms only at the base point,
\begin{equation}
P_{\mathcal C_j}\rightarrow W_j^\dagger P_{\mathcal C_j}W_j .
\end{equation}
The eigenvalues of $P_{\mathcal C_j}$, or its conjugacy class, are therefore gauge invariant.  A pure gauge transformation $\mathcal W_j$ belongs to the IGG only if it commutes with all flux operators based at $j$,
\begin{equation}
\left[\mathcal W_j,P_{\mathcal C_j}\right]=0
\end{equation}
for every closed loop $\mathcal C_j$.

In a canonical gauge the flux operator around a $q$-sided loop has the form
\begin{align}
P_{\mathcal C_j}^{\rm int}
&\propto
(\tau^3)^q
\left(
\rho\, e^{i\phi_{\mathcal C}^{0}\tau^3}
+
\rho'\, e^{i\phi_{\mathcal C}^{1}\tau^3}\tau^1
\right),
\label{eq:integer_flux}\\
P_{\mathcal C_j}^{\rm half}
&\propto
W_j e^{i\phi_{\mathcal C}\tau^3}
(\tau^3)^q W_j^\dagger,
\qquad
W_j\in \mathrm{SU}(2),
\label{eq:odd_flux}
\end{align}
where $P_{\mathcal C_j}^{\rm int}$ and $P_{\mathcal C_j}^{\rm half}$ refer to the integer-spin and half-odd-integer spin cases, respectively.  The phases $\phi_{\mathcal C}^{0}$, $\phi_{\mathcal C}^{1}$, and $\phi_{\mathcal C}$ are the gauge fluxes, or Aharonov--Bohm phases, accumulated by the fermionic spinons around the loop $\mathcal C$.  For integer spin there are two $\mathrm{U}(1)$ phases, one for each of the components $\tau^0$ and $\tau^1$ of the discrete $\mathbb{Z}_2$ sector. Only $\phi^0_{\mathcal C}$, the moduli $\rho,\rho'$, and differences of $\phi^1$ between loops sharing a base point are gauge invariant: under a base-point rotation $e^{i\theta\tau^3}$ the $\tau^1$-sector angle shifts, $e^{-i\theta\tau^3}\big(e^{i\phi^1\tau^3}\tau^1\big)e^{i\theta\tau^3}=e^{i(\phi^1-2\theta)\tau^3}\tau^1$, so the values of $\phi^1$ quoted below are representatives in the canonical gauge of the appendices.  For half-odd-integer spin there is a single $\mathrm{SU}(2)$ flux phase, up to conjugation by $W_j$.

\section{Spin-$1$ PSG classification on the diamond lattice}
\label{sec:diamond_classification}
\begin{figure}
    \includegraphics[width=1.0\linewidth]{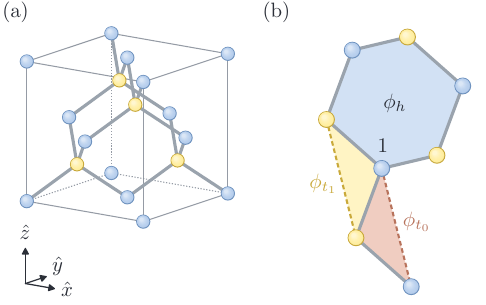}
	\caption{(a) Conventional cubic cell of the diamond lattice, composed of two FCC sublattices (blue and yellow sites). (b) The loops used to characterize the gauge fluxes of the \textit{Ans\"atze} with amplitudes up to second-neighbor bonds: the hexagon ($\phi_h$) and the two triangles ($\phi_{t_0}$, $\phi_{t_1}$), whose second-neighbor bonds are dashed; the base site $(0,0,0,0)$ is labeled `$1$'.}
	\label{fig:lattice_loop}
\end{figure}

We now classify the fully symmetric spin-$1$ mean-field \textit{Ans\"atze} on the diamond lattice, using the lattice conventions and notation of Ref.~\cite{Chauhan-2023}. The diamond lattice [Fig.~\ref{fig:lattice_loop}(a)] has space group $Fd\bar{3}m$ (No.~227). In the coordinates of Ref.~\cite{Chauhan-2023} it is generated by the translations $T_i$, $i=1,2,3$, along the three FCC primitive directions, the screw rotation $S_4$, and the sixfold rotoinversion $\bar C_3$. A fully symmetric \textit{Ansatz} must also preserve time reversal $\mathcal T$ projectively.

We keep amplitudes on first- and second-neighbor bonds only. The classification then gives ten distinct symmetric \textit{Ans\"atze}, listed in Table~\ref{table:ansatz-j1-j2}. The first seven are hopping-only \textit{Ans\"atze}. Their link structure is the same as in the spin-$1/2$ diamond-lattice classification, because the hopping field satisfies
$
\chi_{ji}=\chi_{ij}^\dagger
$
for both integer and half-odd-integer spins. Their IGGs differ, however, because the internal symmetry group of the spin-$1$ representation differs from that of spin-$1/2$ Abrikosov fermions. The two link patterns that give $\mathrm{SU}(2)$ \textit{Ans\"atze} for spin $1/2$ reappear for spin $1$ with IGG $\mathrm{U}(1)\rtimes\mathbb{Z}_2$ (first two rows of Table~\ref{table:ansatz-j1-j2}). The $\mathrm{U}(1)$ hopping \textit{Ans\"atze} of spin $1/2$ also have integer-spin counterparts; with amplitudes up to second neighbors, the seven $\mathrm{U}(1)$ \textit{Ans\"atze} of Ref.~\cite{Chauhan-2023} reduce to the five in rows three to seven.

\begin{table*}
\begin{ruledtabular}
\begin{tabular}{cccc}
\multirow{3}{*}{Label} & 1NN & 2NN & Onsite\\
& \{$u_{OA},u_{OB},u_{OC},u_{OD}$\} & \{$u^0_{O1},u^0_{O2},u^0_{O3},u^0_{O4},u^0_{O5},u^0_{O6},$ & $\{\lambda(0),\lambda(1)\}$  \\
& & \,\,\,\,$u^1_{O1},u^1_{O2},u^1_{O3},u^1_{O4},u^1_{O5},u^1_{O6}\}$ & \\
\hline
$\mathrm{SA3}$& $\{u_1=\chi^{}_1\tau^3\}$ & $\{0\}$ & $\{0\}$ \\
$\mathrm{SB1}$& $\{u_1=\chi^{}_1\tau^3\}$ & $\{0\}$ & $\{0\}$ \\
$\mathrm{UA01}$& $\{u_1=\chi^{}_1\tau^3\}$ & $\{u_2=\chi^{}_2\tau^3\}$ & $\{\lambda\}$ \\
$\mathrm{UA31}$& $\{u_1=\chi^{}_1\tau^3\}$ & $\{u_2=\chi^{}_2\tau^3,u_2,u_2,u_2,u_2,u_2,$ & $\{\lambda,-\lambda\}$ \\
& & $u'_2=-u_2,u'_2,u'_2,u'_2,u'_2,u'_2\}$ & \\
$\mathrm{UB01}$& $\{u_1=\chi^{}_1\tau^3\}$ & $\{u_2=\chi^{}_2\tau^3,-u_2,u_2,u_2,u_2,-u_2,$  & $\{\lambda\}$ \\
& & $u'_2=u_2,u'_2,u'_2,-u'_2,u'_2,u'_2\}$ & \\
$\mathrm{UB31}$& $\{u_1=\chi^{}_1\tau^3\}$ & $\{u_2=\chi^{}_2\tau^3,-u_2,u_2,u_2,u_2,-u_2,$ & $\{\lambda,-\lambda\}$ \\
& & $u'_2=-u_2,u'_2,u'_2,-u'_2,u'_2,u'_2\}$  & \\
$\mathrm{UC20}$& $\{u_1=\chi^{}_1\tau^3\}$ & $\{0\}$ & $\{0\}$ \\
$\mathrm{ZZA30}$& $\{u_1=\chi^{}_1\tau^3+i\Delta^{}_1\tau^1\}$ & $\{u_2=0\}$ & $\{0\}$ \\
$\mathrm{ZZB10}$& $\{u_1=\chi^{}_1\tau^3+i\Delta^{}_1\tau^1\}$ & $\{u_2=0\}$ & $\{0\}$ \\
$\mathrm{ZA3110}$& $\{u_1=\chi^{}_1\tau^3+i\Delta^{}_1\tau^1\}$ & $\{u_2=\chi^{}_2\tau^3\}$ & $\{\lambda\}$ \\
\end{tabular}
\end{ruledtabular}
\caption{Fully symmetric spin-$1$ mean-field \textit{Ans\"atze} on the diamond lattice with amplitudes restricted to first- and second-neighbor bonds. The classification contains two $\mathrm{U}(1)\rtimes\mathbb{Z}_2$ \textit{Ans\"atze}, five $\mathrm{U}(1)$ \textit{Ans\"atze}, two $\mathbb{Z}_2\times\mathbb{Z}_2$ \textit{Ans\"atze}, and one $\mathbb{Z}_2$ \textit{Ansatz}. The first-neighbor bonds are denoted by $u_{OA}$, $u_{OB}$, $u_{OC}$, and $u_{OD}$, while $u^0_{Oi}$ and $u^1_{Oi}$, $i=1,\ldots,6$, denote second-neighbor bonds on the two FCC sublattices. The bond notation and labeling convention follow Ref.~\cite{Chauhan-2023}: for the hopping states the letter (A, B, C) labels the class of translation signs and the digits the remaining PSG parameters, and for the paired states the labels are decoded in Appendices~\ref{app:z2_psW_derivation} and~\ref{app:z2z2_psW_derivation}. Only the link matrix on the representative bond of each shell is listed; the matrices on all symmetry-related bonds are generated from it by the gauge transformations of the corresponding PSG. Rows with identical representative entries (SA3, SB1, UC20; ZZA30, ZZB10) are distinct \textit{Ans\"atze} that differ in their sign structure and gauge fluxes [see Eqs.~\eqref{eq:ZZA30_flux} and \eqref{eq:ZZB10_flux}]. An entry $\{\lambda,-\lambda\}$ denotes a staggered onsite field $\Lambda_i=\pm\lambda\tau^3$ on the two sublattices. For UA31 and UB31 the onsite field is staggered; their symmetric realizations at fixed filling have $n_0+n_1=3$, i.e., $(3/2,3/2)$, $(1,2)$, or $(2,1)$, whereas UA01 and UB01 are symmetric with $n_0=n_1$.} \label{table:ansatz-j1-j2}
\end{table*}

The two classifications differ most once singlet pairing is allowed. For spin $1$ the singlet pairing field is odd under site interchange, while for spin $1/2$ it is even, and this changes which paired \textit{Ans\"atze} are allowed. The spin-$1/2$ diamond-lattice classification has eight short-range $\mathbb{Z}_2$ \textit{Ans\"atze}. For spin $1$, with amplitudes up to second neighbors, only one algebraic $\mathbb{Z}_2$ PSG admits a nonvanishing \textit{Ansatz}, ZA3110 (last row of Table~\ref{table:ansatz-j1-j2}); its derivation is given in Appendix~\ref{app:z2_psW_derivation}. Its self-consistent saddle point has $\chi^{}_1=0$ (Sec.~\ref{sec:j1j2_results}), where the IGG is enlarged to $\mathrm{U}(1)$, so a genuine $\mathbb{Z}_2$ state requires longer-range amplitudes.

In addition, integer-spin systems permit an IGG of the form
$\mathbb{Z}_2\times\mathbb{Z}_2$, which has no analog in
the usual spin-$1/2$ Abrikosov-fermion PSG classification.
This case is obtained by allowing only the $\lambda^0_{ij}$ and
$\lambda^1_{ij}$ components of the link matrix to be nonzero,
thereby breaking the continuous $\mathrm{U}(1)$ part of the
$\mathrm{U}(1)\rtimes\mathbb{Z}_2$ structure down to a discrete subgroup.
The corresponding PSG solutions are given in
Appendix~\ref{app:z2z2_psW_derivation}; they yield two
\textit{Ans\"atze}, labeled $\mathrm{ZZA30}$ and $\mathrm{ZZB10}$, which appear in rows
eight and nine of Table~\ref{table:ansatz-j1-j2}.

We characterize the \textit{Ans\"atze} by their gauge fluxes. Without third-neighbor amplitudes, the loop operators that involve third-neighbor bonds,
\begin{equation}
P_{3t_0}(\phi_{3t_0}),\qquad
P_{3t_1}(\phi_{3t_1}),\qquad
P_{3r}(\phi_{3r}),
\end{equation}
are not defined, and we need only the hexagonal loop $P_h(\phi_h)$ and the two triangular loops $P_{t_0}(\phi_{t_0})$ and $P_{t_1}(\phi_{t_1})$, defined as in Ref.~\cite{Chauhan-2023} and shown in Fig.~\ref{fig:lattice_loop}(b). The \textit{Ans\"atze} are then characterized by the three fluxes
\begin{equation}
(\phi_h,\phi_{t_0},\phi_{t_1}) .
\end{equation}

For integer spin, however, each loop flux has two components,
as described in Eq.~\eqref{eq:integer_flux}.  We denote the flux through a loop $\mathcal C$ by
\begin{equation}
\phi_{\mathcal C}
=
\left(\phi^0_{\mathcal C},\phi^1_{\mathcal C}\right),
\end{equation}
where the two components are associated with the $\tau^0$ and
$\tau^1$ sectors of the integer-spin flux operator.  For
hopping-only states, including the $\mathrm{U}(1)\rtimes\mathbb{Z}_2$
and $\mathrm{U}(1)$ \textit{Ans\"atze}, the coefficient of the $\tau^1$ sector
in Eq.~\eqref{eq:integer_flux} vanishes, $\rho'=0$.  Hence,
each loop carries only the ordinary $\mathrm{U}(1)$ flux component
$\phi^0_{\mathcal C}$.  The flux structures of these
hopping-only states are thus the same as in the
corresponding spin-$1/2$ classification of
Ref.~\cite{Chauhan-2023}.

For paired integer-spin \textit{Ans\"atze} with $\mathbb{Z}_2$ or $\mathbb{Z}_2\times\mathbb{Z}_2$ IGG, both components of the flux must be kept. We write $*$ for a sector that is absent ($\rho=0$ or $\rho'=0$ in Eq.~\eqref{eq:integer_flux}), which is different from a phase equal to zero. For the $\mathbb{Z}_2$ \textit{Ansatz} $\mathrm{ZA3110}$ with the uniform bond pattern of Table~\ref{table:ansatz-j1-j2} and $\chi^{}_1\Delta^{}_1\chi^{}_2\neq0$, the loop products give $u_1u_1^\dagger=(\chi^{2}_1+\Delta^{2}_1)\tau^0+2\chi^{}_1\Delta^{}_1\tau^2$ and $u_1\tau^3u_1^\dagger=(\chi^{2}_1-\Delta^{2}_1)\tau^3$, and hence
\begin{align}
(\phi^0_h,\phi^1_h)&=\left(0,-\frac{\pi}{2}\right),\notag\\
(\phi^0_{t_0},\phi^1_{t_0})&=\left(0,-\frac{\pi}{2}\right),\notag\\
(\phi^0_{t_1},\phi^1_{t_1})&=(0,*).
\label{eq:ZA3110_flux}
\end{align}
At its self-consistent point $\chi^{}_1=0$ (Sec.~\ref{sec:za3110}) all $\tau^1$ sectors vanish and the fluxes reduce to $(\phi^0_h,\phi^0_{t_0},\phi^0_{t_1})=(0,0,\pi)$, the flux pattern of UA31, consistent with the gauge equivalence discussed there. As noted after Eq.~\eqref{eq:odd_flux}, the values of $\phi^1$ are representatives in the canonical gauge.
The two $\mathbb{Z}_2\times\mathbb{Z}_2$ \textit{Ans\"atze}, ZZA30 and ZZB10, have no second-neighbor amplitudes, so their triangular fluxes are not defined and they are characterized by the hexagonal flux alone:
\begin{align}
\mathrm{ZZA30}:\qquad
(\phi^0_h,\phi^1_h)
&=
\left(0,-\frac{\pi}{2}\right),
\label{eq:ZZA30_flux}\\
\mathrm{ZZB10}:\qquad
(\phi^0_h,\phi^1_h)
&=
\left(\pi,\frac{\pi}{2}\right).
\label{eq:ZZB10_flux}
\end{align}

\begin{figure*}[tb]
    \includegraphics[width=1.0\linewidth]{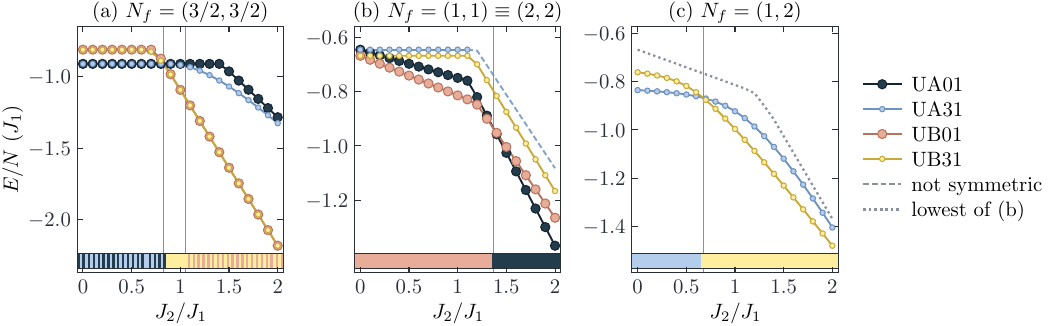}
    \caption{Mean-field energies per site of the second-neighbor hopping \textit{Ans\"atze} of Table~\ref{table:ansatz-j1-j2} for the spin-$1$ $J_1$--$J_2$ diamond-lattice Heisenberg model, for the three implementations of the constraint, labeled by the sublattice fillings $N_f=(n_0,n_1)$; the symmetry-allowed staggered onsite terms of UA31 and UB31 are determined self-consistently. (a) Particle-hole-symmetric implementation, $N_f=(3/2,3/2)$. (b) Uniform particle representation, $N_f=(1,1)$; the hole representation, $(2,2)$, gives an identical diagram (with all second-neighbor amplitudes reversed in sign) because the two are related by the global gauge transformation $\tau^1$. In this sector UA31 and UB31 are symmetric only where their realization is self-dual (see Appendix~\ref{app:selfdual}): UB31 throughout (SB1, and the decoupled class-B state above $J_2/J_1\simeq1.2$) and UA31 for $J_2/J_1\lesssim1.1$ (SA3); the branch of UA31 with $\chi^{}_2\neq0$, which breaks $S_4$ and $\bar C_3$ (its second-neighbor bond energies differ between the two sublattices), is dashed. Particle-hole symmetry is broken explicitly by the constraint for all states in this panel. (c) Particle representation on sublattice $0$ and hole representation on sublattice $1$, $N_f=(1,2)$, the symmetric realization of UA31 and UB31 [$(2,1)$ is its image under $\tau^1$]; UA31$\,(1,2)$ coincides with the $\mathbb{Z}_2$ \textit{Ansatz} ZA3110 at its self-consistent point $\chi^{}_1=0$. The dotted line repeats the lowest energy of (b). The colored strip marks the lowest-energy symmetric \textit{Ansatz} among those computed (striped where two are degenerate), and the vertical lines the transitions between them (the first-neighbor-only states UC20, ZZA30, and ZZB10 are not included; see text).}
    \label{fig:energy_phase_diagram}
\end{figure*}
\section{Mean-field phase diagram of the spin-$1$ $J_1$--$J_2$ model}
\label{sec:j1j2_results}
We now turn to the antiferromagnetic $J_1$--$J_2$ Heisenberg model on the spin-$1$ diamond lattice,
\begin{equation}
H
=
J_1\sum_{\langle ij\rangle_1}\mathbf S_i\cdot\mathbf S_j
+
J_2\sum_{\langle ij\rangle_2}\mathbf S_i\cdot\mathbf S_j ,
\end{equation}
where $\langle ij\rangle_1$ and $\langle ij\rangle_2$ denote first- and second-neighbor bonds. Previous work on this model found a broad nonmagnetic regime for sizable $J_2/J_1$~\cite{Buessen-2018}. We treat it within the zeroth-order fermionic mean-field theory of Sec.~\ref{sec:spinS_formalism}.

The classical model is well understood. The nearest-neighbor ordered state becomes unstable at $J_2/|J_1|=1/8$, and beyond this point the ground states form a two-dimensional spiral surface in momentum space~\cite{Bergman-2007}. Thermal and quantum fluctuations select particular spirals from this surface by order by disorder~\cite{Bergman-2007,Bernier-2008,Chen-2017}, and finite-spin corrections can enlarge nearby collinear regions and stabilize further ordered states near special points of the surface~\cite{Oitmaa-2019,Iqbal-2026}. We ask a complementary question: which fully symmetric fractionalized \textit{Ans\"atze}, among those allowed by the spin-$1$ gauge structure, have the lowest energy in the same frustrated regime?

For spin $1$ the strict particle and hole constraints are
\begin{equation}
{\rm Tr}\!\left[\psi_i\tau^3\psi_i^\dagger\right]
=
\pm 1 ,
\end{equation}
with $+$ for the particle ($N_f=1$) and $-$ for the hole ($N_f=2$) representation. At the saddle-point level we also consider the particle-hole-symmetric implementation,
\begin{equation}
\left\langle
{\rm Tr}\!\left[\psi_i\tau^3\psi_i^\dagger\right]
\right\rangle
=
0 ,
\end{equation}
which corresponds to an average filling $\langle N_f\rangle=3/2$. This is a mean-field treatment of the constraint, not a projection onto either the particle or the hole Hilbert space. For $S=1$ the exchange decouples exactly into the singlet hopping and pairing channels~\cite{Liu-2010a}, which fixes $a(1)=1/2$ in Eq.~\eqref{eq:quadratic_ham}; the resulting quadratic Hamiltonian is Eq.~\eqref{eq:quadratic_ham_mean} of Appendix~\ref{app:zeroth_order}.

We solve the self-consistency equations for the \textit{Ans\"atze} of Table~\ref{table:ansatz-j1-j2} that carry second-neighbor amplitudes (UA01, UA31, UB01, UB31, and ZA3110) and compare their energies per site, $E/N=\langle\hat H^0_{\rm MF}\rangle/N$, as functions of $J_2/J_1$. At the self-consistent point this equals $\langle H\rangle$ in the mean-field state. SA3 and SB1 are the $\chi^{}_2=\lambda=0$ limits of UA and UB class respectively and appear in their scans. UC20, ZZA30, and ZZB10 have first-neighbor amplitudes only, so their energies do not depend on $J_2$; we have not computed them. Since such a state could still be lowest over a range of small $J_2$, ``lowest state'' below always refers to the five families with second-neighbor amplitudes together with SA3 and SB1.

Two features of the integer-spin gauge structure matter for a correct comparison. The first concerns UA31 and UB31, whose symmetry-allowed onsite term is staggered, $\Lambda_i=\pm\lambda\tau^3$. This term must be determined self-consistently. The Bloch Hamiltonian of these states is traceless at every $\mathbf k$, so the total filling is correct for any $\lambda$, but $\lambda$ alone controls how the fermions are shared between the two sublattices. With $\lambda=0$ the total is right while the constraint on each site is violated; for UA31 at $N_f=3/2$ and $J_2/J_1=2$ the violation is $|\langle n_\mu\rangle-3/2|=0.33$, and the energy is spuriously lowered by $0.11\,J_1$.

The second feature is the global gauge transformation $\tau^1$. It maps $\chi^3\to-\chi^3$, $\lambda\to-\lambda$, and $\zeta\to-\zeta$, and hence maps the particle implementation exactly onto the hole implementation. The $N_f=1$ and $N_f=2$ phase diagrams are identical, with all second-neighbor hopping amplitudes reversed in sign. The same transformation applied to one sublattice only relates the paired \textit{Ansatz} ZA3110 to UA31: the two describe the same state, with the constraint sectors exchanged on that sublattice.

The transformation $\tau^1$ also decides which constraint sectors are symmetric. For UA31 and UB31 the staggered onsite term is allowed only because the gauge parts of the sublattice-exchanging operations $S_4$ and $\bar C_3$ lie in the $\mathbb{Z}_2$ coset. These operations leave the staggered field invariant but reverse the uniform chemical potential that fixes the overall filling, so they map the sector $(n_0,n_1)$ onto $(3-n_1,3-n_0)$ (Appendix~\ref{app:selfdual}). At $N_f=1$ the symmetric realization of either \textit{Ansatz} therefore has the two sublattices in opposite constraint sectors, $(n_0,n_1)=(1,2)$; for UA31 this is ZA3110 with $\chi^{}_1=0$ (uniform $\zeta=+1$ and singlet pairing on the first-neighbor bonds). A realization with $\langle n_i\rangle=1$ on both sublattices is symmetric only if it is self-dual, i.e., if it coincides with the $(1,1)$ realization with reversed $\chi^{}_2$ and $\lambda$; Sec.~\ref{sec:uniform} and Appendix~\ref{app:selfdual} discuss when this holds. For UA01 and UB01, whose onsite term is uniform, the gauge parts of $S_4$ and $\bar C_3$ lie in $\mathrm{U}(1)$, and only the uniform realization $(1,1)$ is symmetric.

The mixed-sector realizations must be included in the particle and hole phase diagrams alongside the uniform ones. We label the three implementations by the sublattice fillings $N_f=(n_0,n_1)$: $(3/2,3/2)$, $(1,1)\equiv(2,2)$, and $(1,2)\equiv(2,1)$. In the mixed sector $(1,2)$ the particle representation is imposed on sublattice $0$ and the hole representation on sublattice $1$, $\langle{\rm Tr}[\psi_i\tau^3\psi_i^\dagger]\rangle=\pm1$. The average filling is $3/2$, as in the particle-hole-symmetric treatment, but every site has the integer filling of a spin-$1$ sector, as in the uniform sector. The staggered multiplier enforces $n_0-n_1=-1$ at every self-consistency step, and its contribution $\lambda(n_0-n_1)/2$ per site is subtracted from the reported energies. A uniform multiplier, as in UA01 and UB01, couples only to the total density, which is already fixed, and cannot produce this imbalance; so these two \textit{Ans\"atze} have no $(1,2)$ realization, in agreement with the symmetry argument. Figure~\ref{fig:energy_phase_diagram} shows the energies of the three implementations.
\subsection{Particle--hole symmetric constraint, $N_f=(3/2,3/2)$}
\label{sec:phsym}
Figure~\ref{fig:energy_phase_diagram}(a) shows the energies. At this filling the constraint is invariant under the global transformation $\tau^1$, which then acts as a gauge transformation. It maps $(\chi^{}_2,\lambda)\to(-\chi^{}_2,-\lambda)$ and leaves the first-neighbor amplitudes unchanged up to a sublattice sign in $\mathrm{U}(1)$. The point $\chi^{}_2=\lambda=0$, where this element belongs to the IGG, is therefore always a self-consistent solution, and for all four hopping \textit{Ans\"atze} it is the stable one at small $J_2/J_1$. There the second-neighbor amplitude is exactly zero, and the energies of UA01, UA31, and ZA3110 equal that of the parent state SA3, $E/N=-0.9104\,J_1$, while UB01 and UB31 equal SB1, $E/N=-0.8111\,J_1$.

This regime ends in one of two ways. In the two states with a staggered onsite term, $(\chi^{}_2,\lambda)$ turns on continuously: at $J_2/J_1=0.840$ in UA31 (equivalently ZA3110), whose nodal spectrum then becomes gapped, and at $J_2/J_1\simeq0.62$ in UB31. The two signs of $(\chi^{}_2,\lambda)$ are related by $\tau^1$ and describe the same physical state, so this is not a symmetry breaking. The IGG is reduced from $\mathrm{U}(1)\rtimes\mathbb{Z}_2$ to $\mathrm{U}(1)$, the lattice symmetries and time reversal remain projectively realized, and the staggered $\lambda$ keeps both sublattices at $\langle\hat N_i\rangle=3/2$. In the sense of the constraint, particle-hole symmetry is preserved throughout panel (a), whereas the constraints of panels (b) and (c) break it explicitly.

In UA01 and UB01, which have no staggered term, the first-neighbor amplitude instead collapses discontinuously, $\chi^{}_1\to0$, at $J_2/J_1\simeq0.74$ (UB01) and $1.42$ (UA01). The two FCC sublattices then decouple, and the energy becomes exactly linear in $J_2$, $E/N=-6\,\chi^{2}_2\,J_2$, with a $J_2$-independent $\chi^{}_2$ ($0.4266$ for UB01, $0.3270$ for UA01). In UB31 the first-neighbor amplitude decreases continuously and vanishes at $J_2/J_1\simeq1.0$. Beyond it UB31 coincides with the decoupled UB01 state: once the sublattices decouple, the second-neighbor patterns of the two states on sublattice $\mu=1$ differ only by an overall sign, which does not matter at half filling of a single sublattice, and the staggered $\lambda$ vanishes.

For $0.62\lesssim J_2/J_1\lesssim1.0$ the coupled UB31 state with $(\chi^{}_2,\lambda)\neq0$ lies below the decoupled state, by $0.006\,J_1$ at $J_2/J_1=0.85$ and $0.002\,J_1$ at $0.9$, and the difference vanishes at $J_2/J_1\simeq1.0$. The coupled state is gapped; the gap decreases from $0.94\,J_1$ at $J_2/J_1=0.8$ to $0.49\,J_1$ at $0.9$ and closes as $\chi^{}_1\to0$. Among the computed states, the phase diagram is
\begin{align}
0\le J_2/J_1\lesssim 0.82:&\quad \mathrm{SA3}\ (\text{UA01}=\text{UA31}),\\
0.82\lesssim J_2/J_1\lesssim 1.0:&\quad \text{UB31 (coupled)},\\
J_2/J_1\gtrsim 1.0:&\quad \text{UB31}=\text{UB01 (decoupled)},
\end{align}
where SA3 is realized by UA01, UA31, and ZA3110 alike.

\subsection{Uniform particle and hole constraints, $N_f=(1,1)\equiv(2,2)$}
\label{sec:uniform}
Figure~\ref{fig:energy_phase_diagram}(b) shows the $(1,1)$ energies; the $(2,2)$ diagram is identical. In this sector the constraint breaks particle-hole symmetry explicitly for all four states. The breaking of $S_4$ and $\bar C_3$ discussed below has a different origin: the gauge part $\tau^1$ of these operations maps the $(1,1)$ sector onto $(2,2)$, so a $(1,1)$ realization of UA31 or UB31 is symmetric only if it is self-dual (Appendix~\ref{app:selfdual}).

At $J_2=0$ the hopping states have $E/N=-0.6458\,J_1$ (SA3) and $-0.6681\,J_1$ (SB1). How they evolve with $J_2$ follows from an identity of the diamond lattice, $\Gamma_2(\mathbf k)=|g_1(\mathbf k)|^2-4$, with $g_1=\sum_{\rm 1NN}e^{i\mathbf k\cdot\bm\delta}$ and $\Gamma_2=\sum_{\rm 2NN}e^{i\mathbf k\cdot\mathbf d}$. For UA01 both the first- and second-neighbor dispersions depend on $\mathbf k$ only through $|g_1|$. The set of occupied states, and hence $\chi^{}_1$ and $\chi^{}_2$, do not change with $J_2$ until the band ordering changes, and between such rearrangements the energy is exactly linear in $J_2$ ($E/N=-0.6459-0.104\,J_2$ for UA01 up to $J_2/J_1=1.1$, then $-0.8205$ at $1.2$ with a much larger $\chi^{}_2$). Among the symmetric states of this sector, UB01 is lowest up to $J_2/J_1\simeq1.37$ and UA01 above, in a form with a nearly vanishing first-neighbor amplitude ($E/N=-1.3675\,J_1$ at $J_2/J_1=2$).

The $(1,1)$ realizations of UA31 and UB31 converge to the nearest-neighbor states SA3 and SB1 ($\chi^{}_2=0$) for $J_2/J_1\lesssim1.1$. Above this, UB31 becomes the decoupled class-B state, which is self-dual and hence symmetric. UA31 instead develops $\chi^{}_2\neq0$ and a staggered $\lambda$, and it breaks the sublattice-exchanging operations $S_4$ and $\bar C_3$. With $\chi^{}_1\simeq0$ its two FCC sublattices decouple and carry second-neighbor hopping of opposite sign at filling $1/3$. Because the FCC lattice is not bipartite, these two sublattice states are physically different: their second-neighbor bond energies differ by about $25\%$, although $S_4$ maps one sublattice onto the other. Both states lie above UA01 and UB01 throughout.

\subsection{Mixed constraint, $N_f=(1,2)$}
\label{sec:mixed}
Figure~\ref{fig:energy_phase_diagram}(c) shows the energies of the symmetric mixed-sector realizations of UA31 and UB31. ZA3110 $\equiv$ UA31$\,(1,2)$ has $E/N=-0.8355\,J_1$ at $J_2=0$ and is the lowest computed state up to $J_2/J_1\simeq0.67$. The mixed-sector UB31 state starts higher, at $-0.7616\,J_1$, but gains energy faster with $J_2$. It is the lowest computed state above $J_2/J_1\simeq0.67$. Unlike the class-B states at $N_f=3/2$, it keeps a sizable first-neighbor amplitude ($\chi^{}_1=0.48$ at $J_2/J_1=2$) and is gapped throughout, with a spinon gap that grows from $1.81\,J_1$ at $J_2/J_1=0.7$ to $2.45\,J_1$ at $J_2/J_1=2$. ZA3110 is also gapped, with a largest gap $2E_{\min}=2.16\,J_1$ at $J_2/J_1\simeq1.1$.

The mixed sector lies below the uniform one at every coupling [dotted line in Fig.~\ref{fig:energy_phase_diagram}(c)]. Both sectors give every site the integer filling of a spin-$1$ sector, so the lowest computed state in the particle and hole representations is ZA3110 $\equiv$ UA31$\,(1,2)$ for $J_2/J_1\lesssim0.674$ and the mixed-sector UB31 state above.

\subsection{The $\mathbb{Z}_2$ \textit{Ansatz} ZA3110}
\label{sec:za3110}
When all three amplitudes of ZA3110 are left free, the self-consistent first-neighbor hopping $\chi^{}_1$ vanishes at every point we examined, for all three constraints. The surviving state, $u_1=i\Delta^{}_1\tau^1$, $u_2=\chi^{}_2\tau^3$ has an enlarged IGG $\mathrm{U}(1)$, and is thus a
$\mathrm{U}(1)$ state written in a pairing gauge rather than a genuine
$\mathbb{Z}_2$ state at the saddle point. This is a gauge image of UA31 described above in a modified constraint sector discussed in the following.

The connection is made by the gauge transformation $W=\tau^0$ on one sublattice
and $W=e^{i\frac{\pi}{2}\tau^3}\tau^1$ on the other. It maps $i\Delta^{}_1\tau^1$ onto the
hopping $\Delta^{}_1\tau^3$, so that $\Delta^{}_1$ becomes the first-neighbor hopping of UA31;
it turns the uniform $\chi^{}_2$ and $\lambda$ into the staggered $\pm \chi^{}_2$ and
$\pm\lambda$; and, because $W$ contains $\tau^1$, it sends $N_f\to3-N_f$ on the
rotated sublattice. Hence ZA3110 at $N_f=3/2$ is UA31 at $(3/2,3/2)$, contained
in Fig.~\ref{fig:energy_phase_diagram}(a), and ZA3110 at $N_f=1$ or $2$ is UA31
at $(1,2)$ or $(2,1)$, contained in Fig.~\ref{fig:energy_phase_diagram}(c).
Up to second neighbors the classification contains no realized
$\mathbb{Z}_2$ state: the algebraic $\mathbb{Z}_2$ PSG ZA3110 yields a
$\mathrm{U}(1)$ state at the saddle point, adds no saddle point beyond the
$\mathrm{U}(1)$ hopping states, and is not drawn separately. This is specific to
$S>1/2$: for $S=1/2$ the particle and hole representations coincide, and the
analogous rotation does not change the constraint sector.

\subsection{Particle-hole-averaged representation}
\label{sec:lzn}
Since the phase diagram depends on how the constraint is implemented, one may ask whether there is a representation in which any generic $\mathrm{U}(1)\rtimes\mathbb{Z}_2$ gauge transformation does not change the constraint sector at all. For integer spin, Liu, Zhou, and Ng~\cite{Liu-2010a} constructed one by averaging the partition function over the particle and hole constraints on every site, $\prod_i\tfrac12[\delta(\hat N_i-1)+\delta(\hat N_i-2S)]$. This representation of the spin model is exact, and the constraint enters the Lagrangian through the weight $\cosh[(2S-1)\tilde\lambda_i/2]$ of the multiplier $\tilde\lambda_i$. The weight is even under $\tilde\lambda_i\to-\tilde\lambda_i$, which is the action of the $\tau^1\in\mathbb{Z}_2$ coset of $\mathrm{U}(1)\rtimes\mathbb{Z}_2$, so the Lagrangian is invariant under the full internal group $\mathrm{U}(1)\rtimes\mathbb{Z}_2$. The constraint then places no restriction on the PSG: $\tilde\lambda_i$ is an onsite field like any other, uniform for UA01 and UB01 and staggered for UA31 and UB31, and every saddle point is symmetric. The local rotation that maps ZA3110 onto UA31 leaves the averaged constraint invariant, so these two \textit{Ans\"atze} again have the same saddle points.

\begin{figure*}[t]
    \includegraphics[width=1.0\linewidth]{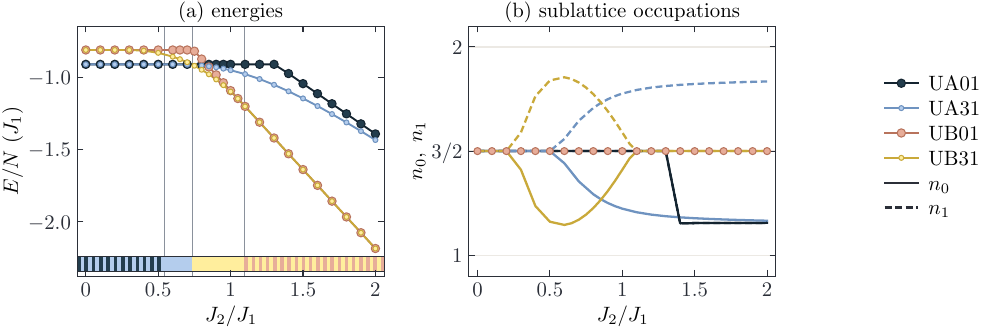}
    \caption{Zero-temperature saddle points ($\beta\to\infty$) of the particle-hole-averaged representation~\cite{Liu-2010a} for the hopping \textit{Ans\"atze}; ZA3110 coincides with UA31. (a) Mean-field energies per site, with the lowest state marked as in Fig.~\ref{fig:energy_phase_diagram}; the limit $\beta\to0$ is Fig.~\ref{fig:energy_phase_diagram}(a). (b) Sublattice occupations $n_0$ (solid) and $n_1$ (dashed), $n_\mu=\langle\hat N_i\rangle$ for $i$ in sublattice $\mu$: UA31 and UB31 (staggered multiplier, $n_0+n_1=3$), UA01 (uniform multiplier, $n_0=n_1$), and UB01 (circles), which has $n_0=n_1=3/2$ throughout.}
    \label{fig:lzn}
\end{figure*}

At the saddle point the mean-field Hamiltonian acquires the terms $\sum_i[\tilde\lambda_i(\hat N_i-\tfrac{2S+1}{2})-\beta^{-1}\ln\cosh\tfrac{\beta(2S-1)\tilde\lambda_i}{2}]$, and the constraint becomes
\begin{equation}
\langle\hat N_i\rangle-\frac{2S+1}{2}=\frac{2S-1}{2}\tanh\frac{\beta(2S-1)\tilde\lambda_i}{2},
\label{eq:lzn}
\end{equation}
where $\beta$ is the imaginary-time interval on which the average over the two constraints is inserted. Inserting it on every time slice and taking the continuum limit gives $\beta\to0$; a single insertion at temperature $T$ gives $\beta=1/T$. The unscaled form written in Ref.~\cite{Liu-2010a} corresponds to $\beta=1/J$. For fixed amplitudes, Eq.~\eqref{eq:lzn} is the maximum of a concave function of $\tilde\lambda_i$ and has a unique solution. For $\beta\to0$ it reduces to $\langle\hat N_i\rangle=(2S+1)/2$, the particle-hole-symmetric implementation of Fig.~\ref{fig:energy_phase_diagram}(a). For $\beta\to\infty$ it relaxes the constraint to the inequality $1\le\langle\hat N_i\rangle\le2S$, with $\tilde\lambda_i=0$ whenever the inequality holds; the exact sectors $(1,1)$ and $(1,2)$ are then saddle points only if the band structure itself produces integer occupations.

Figure~\ref{fig:lzn} shows the zero-temperature saddle points of a single insertion, $\beta\to\infty$; the opposite limit, $\beta\to0$, is Fig.~\ref{fig:energy_phase_diagram}(a). States with particle-hole-symmetric spectra are the same in both limits: SA3, SB1, and the decoupled class-B state. The limits differ only where the single-particle spectrum is not particle-hole symmetric, so that without the multiplier the occupations move away from $3/2$. UA31 and UB31 then develop occupations with $\langle\hat N_0\rangle+\langle\hat N_1\rangle=3$, and the decoupled form of UA01 develops uniform occupations $\langle\hat N_0\rangle=\langle\hat N_1\rangle\neq3/2$. In UB31, particle-hole symmetry in the sense $\langle\hat N_i\rangle\neq3/2$ is broken from $J_2/J_1\simeq0.25$; for $\beta\to0$ it is never broken, and $(\chi^{}_2,\lambda)$ sets in only at $0.62$. The occupations move towards the exact sectors, reaching $(\langle\hat N_0\rangle,\langle\hat N_1\rangle)=(1.15,1.85)$ for UB31 at $J_2/J_1=0.6$ and $(1.17,1.83)$ for UA31 at $J_2/J_1=2$. The averaged representation thus pushes the staggered \textit{Ans\"atze} towards the mixed sector $(1,2)$, which is lowest in Fig.~\ref{fig:energy_phase_diagram}(c).

The lowest states come in the same order as in Fig.~\ref{fig:energy_phase_diagram}(a): SA3, then a state with $(\chi^{}_2,\lambda)\neq0$, then the decoupled class-B state. The middle window is wider, with UA31 lowest for $0.55\lesssim J_2/J_1\lesssim0.74$ and UB31 up to $J_2/J_1\simeq1.1$, compared with UB31 alone for $0.82\lesssim J_2/J_1\lesssim1.0$ when $\beta\to0$. The unscaled form of Ref.~\cite{Liu-2010a}, $\beta=1/J_1$, lies between the two limits; its first boundary is at $J_2/J_1\simeq0.79$. Because the averaged representation enforces the constraint only on average, its energies lie below those of the exact sectors, as in the one-dimensional case~\cite{Liu-2010a}. It gives a manifestly symmetric mean-field framework, but for the spin model the relevant comparison is between the exact sectors of Secs.~\ref{sec:uniform} and \ref{sec:mixed}.

\subsection{Comparison with previous studies and experiment}
\label{sec:comparison}
The pseudofermion functional renormalization group study of Ref.~\cite{Buessen-2018} found that for $S\le1$ the $J_1$--$J_2$ diamond antiferromagnet does not order in the spiral regime $J_2/J_1>1/8$ and forms a quantum spiral spin liquid, whereas for $S\ge3/2$ it orders. In our particle-hole-symmetric phase diagram, the region $J_2/J_1\gtrsim0.8$ of that regime belongs to the class-B states, which for $J_2/J_1\gtrsim1.0$ reduce at the saddle point to decoupled FCC sublattices with the class-B second-neighbor hopping pattern. In the particle and hole representations the same region belongs to the mixed-sector UB31 state. This state is gapped, with a spinon gap of about $2\,J_1$, and so is ZA3110 at smaller $J_2/J_1$. The decoupled class-B state of the particle-hole-symmetric treatment is instead gapless: its bands touch at $\mathbf k=(\pi/2,\pi/2,\pi/2)$ and the equivalent points of the reduced zone of the 8-site cell, and the gap of the coupled UB31 state that precedes it ($0.49\,J_1$ at $J_2/J_1=0.9$) closes as $\chi^{}_1\to0$.

The spin-$1$ diamond-lattice material NiRh$_2$O$_4$ shows no magnetic order and has gapped magnetic excitations in inelastic neutron scattering~\cite{Chamorro-2018}. Gapped spinons are a feature of the candidate states of the particle and hole representations in this regime, but the comparison can only be qualitative. The neutron gap belongs to a physical spin excitation; the measured sample was substoichiometric and tetragonally distorted, with a sizable single-ion anisotropy~\cite{Chamorro-2018,Buessen-2018}; and a quantitative comparison requires projection and gauge fluctuations beyond the present saddle-point treatment.
\section{General filling constraint on gapped $\mathrm{U}(1)$ states}
\label{sec:filling}
\begin{figure*}[t]
    \includegraphics[width=1.0\linewidth]{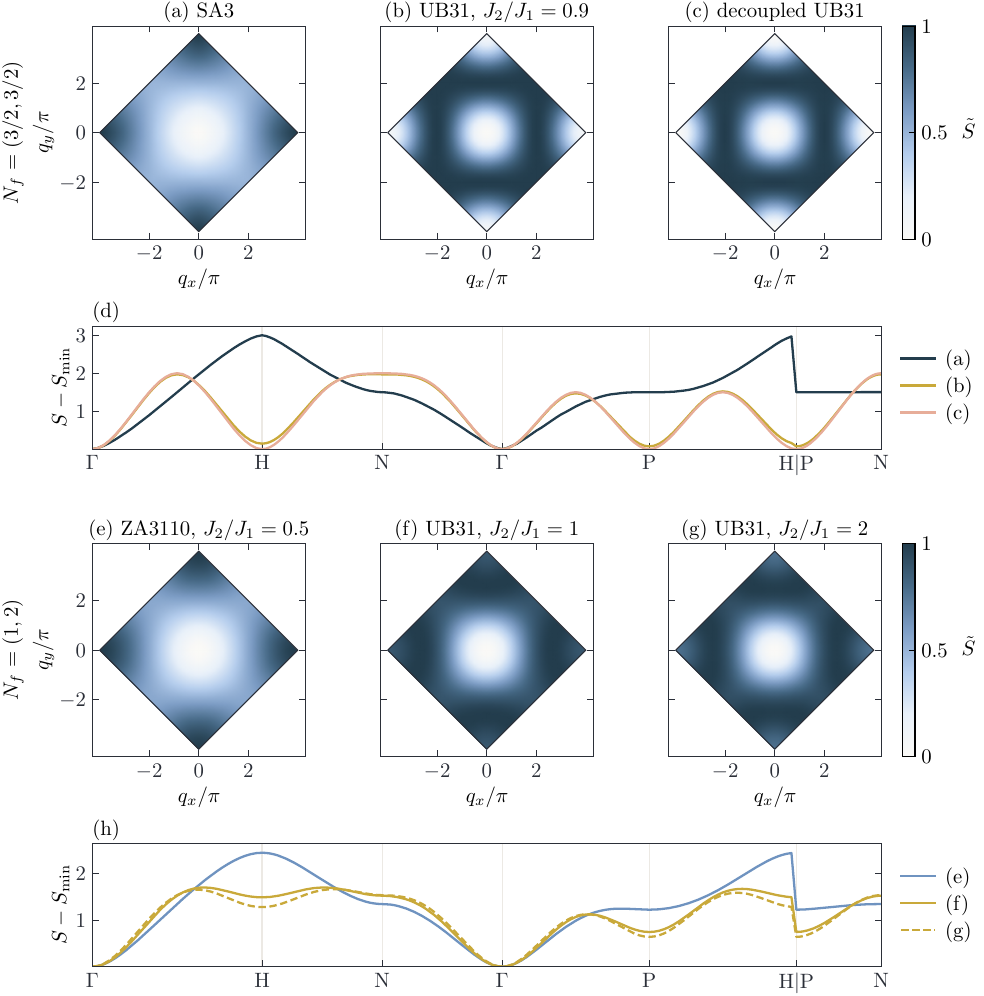}
    \caption{Mean-field static structure factor $S(\mathbf q)$ of the lowest computed state in each regime, evaluated from the particle-hole bubble of the self-consistent states. (a)--(c) $N_f=(3/2,3/2)$: SA3 ($J_2/J_1<0.82$), the coupled UB31 state at $J_2/J_1=0.9$, and the decoupled class-B state, UB31 $=$ UB01 ($J_2/J_1\gtrsim1.0$). (e)--(g) $N_f=(1,2)$: ZA3110 $\equiv$ UA31$\,(1,2)$ at $J_2/J_1=0.5$ and UB31$\,(1,2)$ at $J_2/J_1=1$ and $2$. The panels show the $q_z=0$ plane of the extended Brillouin zone, normalized for each state, $\tilde S=(S-S_{\min})/(S_{\max}-S_{\min})$. (d), (h) $S-S_{\min}$ along the path $\Gamma$--H--N--$\Gamma$--P--H$|$P--N of the extended zone [H$=(0,0,4\pi)$, N$=(0,2\pi,2\pi)$, P$=(2\pi,2\pi,2\pi)$] for the three states of the row; colors follow Fig.~\ref{fig:energy_phase_diagram}.}
    \label{fig:ssf_planes}
\end{figure*}

\begin{figure*}[t]
    \includegraphics[width=1.0\linewidth]{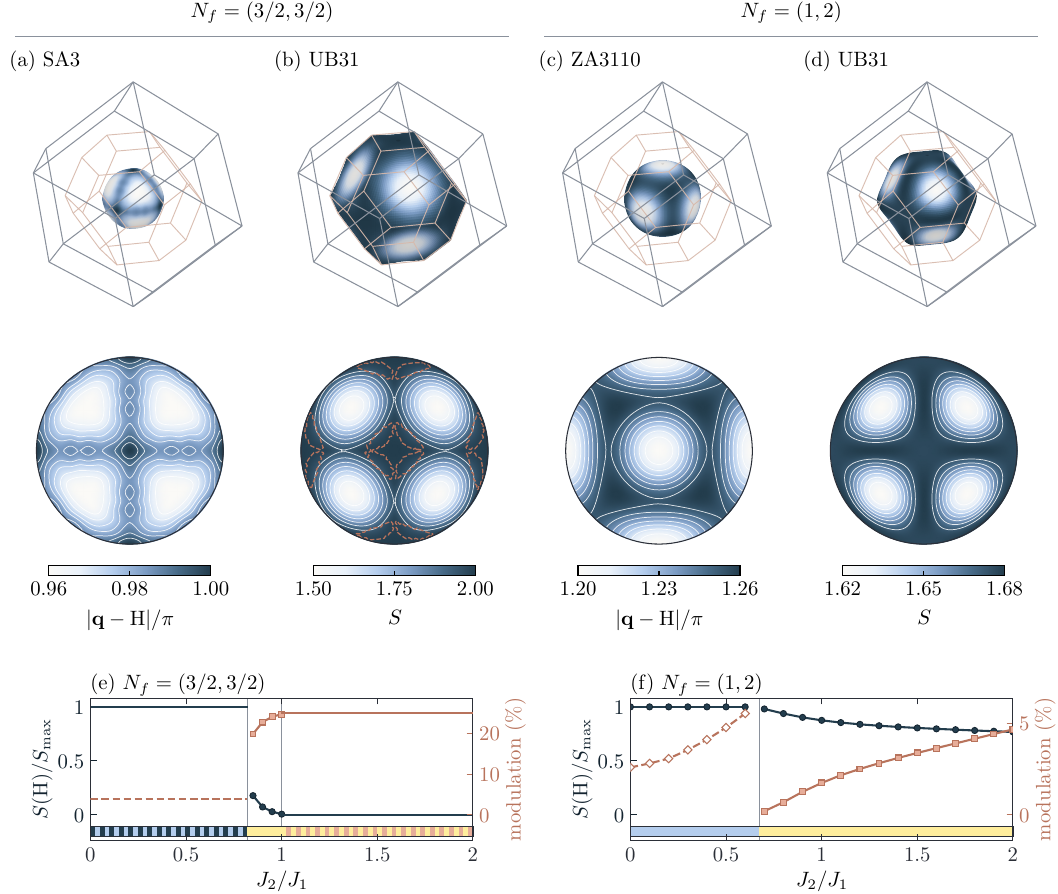}
    \caption{Equal-time structure factor $S(\mathbf q)$ in the extended Brillouin zone: one representative for each qualitatively distinct $S(\mathbf q)$ of the lowest computed state, (a), (b) for $N_f=(3/2,3/2)$ [SA3, and the class-B state, shown in its decoupled form] and (c), (d) for $N_f=(1,2)$ [ZA3110 at $J_2/J_1=0.5$ and UB31 at $J_2/J_1=1.5$]. Top row: three-dimensional representation in the extended zone centered at $\mathrm{H}=(4\pi,0,0)$ (gray rhombic dodecahedron, whose fourfold and threefold vertices are $\Gamma$ and P points), with the first Brillouin zone centered at H (thin lines). For the states with a maximum at H, (a) and (c), the surface $S=0.8\,S_{\max}$ is shown, colored by its distance from H; for the shell states, (b) and (d), the ridge of maxima of $S$ along every direction from H (within the first zone around H) is shown, colored by the intensity on the ridge. Middle row: the same quantity on an equal-area projection of the directions from H (hemisphere about $[001]$, $[100]$ to the right); the dashed lines in (b) enclose the directions in which the ridge lies on the first-zone boundary. Each panel has its own color scale. (e), (f) $S(\mathrm H)/S_{\max}$ (left axis) and the peak anisotropy (open diamonds) or ridge modulation (squares) (right axis) of the lowest computed state versus $J_2/J_1$, with the lowest state marked as in Fig.~\ref{fig:energy_phase_diagram}.}
    \label{fig:ssf_ebz_self}
\end{figure*}

Why the uniform sector is gapless while the mixed sector is gapped follows from a counting argument that holds for any spin $S$ and any lattice. A $\mathrm{U}(1)$ \textit{Ansatz} conserves the spinon number in a suitable gauge, and its $2S+1$ flavors occupy identical bands. Let the unit cell contain $n_0$ sites, $n_p$ of them in the particle sector ($N_i=1$) and $n_h$ in the hole sector ($N_i=2S$); these are the only two sectors that carry spin $S$. Let the projective translations enlarge the unit cell of the gauge-fixed \textit{Ansatz} by a factor $q$. For the time-reversal-symmetric \textit{Ans\"atze} considered here the translation phases are $\pm1$, and $q$ is a power of two ($q=1$ for class A and $q=4$ for class B). The number of occupied states per flavor and enlarged unit cell is then
\begin{equation}
F=q\,\frac{n_p+2S\,n_h}{2S+1}.
\label{eq:filling}
\end{equation}
A gapped mean-field state needs an integer number of filled bands, $F\in\mathbb{Z}$. Otherwise a band is partially filled with a fractional Luttinger volume, and the state has a spinon Fermi surface. The condition is necessary but not sufficient, as the nodal SA3 state at $F=1$ shows.

In the uniform sectors $F=qn_0/(2S+1)$ or $qn_0\,2S/(2S+1)$. Since $2S$ and $2S+1$ are coprime, a gap requires $(2S+1)\mid qn_0$, which for integer $S$ and time-reversal-symmetric \textit{Ans\"atze} means $(2S+1)\mid n_0$. For $S=1$ on the diamond lattice ($n_0=2$) the particle sector has $F=2/3$ for the class-A and $F=8/3$ for the class-B \textit{Ans\"atze}. Every symmetric $\mathrm{U}(1)$ state of the uniform sectors $(1,1)$ and $(2,2)$ therefore has a spinon Fermi surface, as we find numerically. In the mixed sector, $2S\equiv-1\pmod{2S+1}$ gives $F\in\mathbb{Z}$ whenever $(2S+1)\mid q(n_p-n_h)$, in particular for every $S$ when the two sublattices of a bipartite lattice are in opposite sectors, $n_p=n_h$. Paired ($\mathbb{Z}_2$ and $\mathbb{Z}_2\times\mathbb{Z}_2$) states do not conserve the spinon number and are not constrained.

A gapped symmetric mean-field state of an integer-spin magnet with $(2S+1)\nmid n_0$ must therefore be paired, or gauge equivalent to a mixed-sector $\mathrm{U}(1)$ state, as ZA3110 is. This is the free-parton analog of the flux-insertion argument for fermions at fractional filling~\cite{Oshikawa-2000}. It does not constrain the spin model itself: the Lieb-Schultz-Mattis theorem~\cite{LSM-1961,Hastings-2004} restricts only the spin per unit cell, which is an integer here, and beyond mean field the spinon number is a gauge charge.

The same counting decides the fate of the nodal structure inherited from the spin-$1/2$ classification. The nodal manifold of the spin-$1/2$ SA3 state, protected by the projective screw and rotoinversion symmetries~\cite{Chauhan-2023}, is also present in the corresponding spin-$1$ hopping \textit{Ansatz}. Whether it lies at the Fermi level depends on how the constraint is implemented. In the hopping form of these states the spinon filling is $1/3$ in the particle representation and $2/3$ in the hole representation, and the nodal manifold is replaced by a Fermi surface. It lies at the Fermi energy only at the particle-hole-symmetric saddle point, where the filling is $1/2$. The mixed-sector states that are lowest in the particle and hole representations, ZA3110 and UB31, are fully gapped.

\section{Static structure factor}
\label{sec:ssf}

\begin{figure*}[t]
    \includegraphics[width=1.0\linewidth]{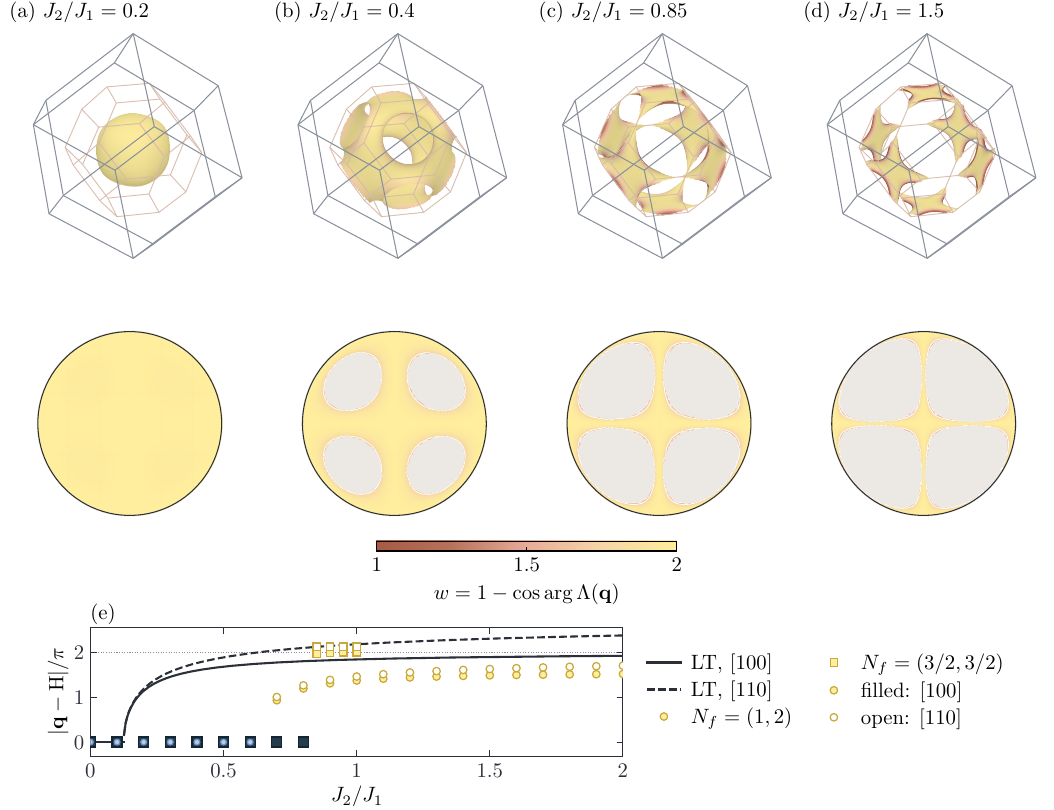}
    \caption{Comparison with the classical Luttinger-Tisza analysis, in the frame of Fig.~\ref{fig:ssf_ebz_self}. (a)--(d) The classical spiral surface $|\Lambda(\mathbf q)|=J_1/2J_2$ inside the first zone centered at H for $J_2/J_1=0.2$, $0.4$, $0.85$ (the values of Fig.~2 of Ref.~\cite{Bergman-2007}, whose f.c.c.\ zone centered at $\Gamma$ in the ferromagnetic gauge $J_1\to-J_1$ corresponds to this zone) and $1.5$, colored by its weight $w=1-\cos\arg\Lambda(\mathbf q)$ in the true-position structure factor, i.e., the $T\to0$ classical intensity along the surface. Middle row: $w$ on the first crossing of the surface along each direction from H (equal-area projection as in Fig.~\ref{fig:ssf_ebz_self}; gray: no crossing, the surface is open along $\langle111\rangle$). (e) Distance from H of the classical surface (lines) and of the maxima of the lowest mean-field states (symbols; zero for the states with a maximum at H) along $[100]$ (filled) and $[110]$ (open), for $N_f=(1,2)$ (circles) and $N_f=(3/2,3/2)$ (squares), with colors as in Fig.~\ref{fig:energy_phase_diagram}. Classically, thermal order by disorder selects wave vectors along $(q,q,q)$ for $1/8<J_2/J_1<1/4$, around the holes for $1/4<J_2/J_1<1/2$, along $(q,q,0)$ for $1/2<J_2/J_1<2/3$, and around $(q,0,0)$ for larger $J_2/J_1$~\cite{Bergman-2007}.}
    \label{fig:lt_comparison}
\end{figure*}

Figures~\ref{fig:ssf_planes} and \ref{fig:ssf_ebz_self} show the equal-time structure factors of the lowest computed states. In every regime the intensity is organized around the N\'eel wave vector $\mathrm{H}=(4\pi,0,0)$. Scanning $J_2/J_1$ gives four qualitatively different forms [Fig.~\ref{fig:ssf_ebz_self}(e,f)], and within each regime $S(\mathbf q)$ changes continuously. At $N_f=3/2$, SA3 has a single maximum at H, elongated along $\langle100\rangle$. The class-B states have $S(\mathrm H)\to0$ and a shell of maxima on the boundary of the first zone around H, strongest at the X points and with deep minima along $\langle111\rangle$ ($25\%$ modulation); the coupled UB31 state differs from the decoupled one only by a small $S(\mathrm H)$, which vanishes continuously. At $N_f=1,2$, ZA3110 also has its maximum at H, but with the opposite anisotropy, flattened along $\langle100\rangle$. The mixed-sector UB31 state has a closed shell of maxima inside the zone on top of a broad maximum at H, with minima along $\langle111\rangle$ whose depth grows from $0.2\%$ at $J_2/J_1=0.7$ to $5\%$ at $J_2/J_1=2$.

Figure~\ref{fig:lt_comparison} compares these results with the classical Luttinger-Tisza (LT) analysis~\cite{Bergman-2007}. In the true-position (atomic) gauge the LT matrix has eigenvalues $\omega_\pm(\mathbf q)=J_2(|g_1(\mathbf q)|^2-4)\pm J_1|g_1(\mathbf q)|$. For $J_2/J_1>1/8$ the minima form the spiral surface $|g_1(\mathbf q)|=J_1/2J_2$. In this gauge the surface is centered at H; it is closed for $J_2/J_1<1/4$ and open along $\langle111\rangle$ above. In the structure factor the lower LT mode carries the weight $w(\mathbf q)=1-\cos\arg g_1(\mathbf q)$, which equals $2$ at H and lies between $1$ and $2$ on the surface within the first zone around H.

Classically, spiral-surface correlations set in at $J_2/J_1=1/8$. The lowest mean-field states, by contrast, keep their maximum at H up to $J_2/J_1=0.82$ ($N_f=3/2$) and $0.674$ ($N_f=1,2$) [Fig.~\ref{fig:lt_comparison}(e)]. Above these values the shells of both class-B states have the topology of the open classical surface, with minima along $\langle111\rangle$. The shell of the $N_f=3/2$ states lies close to the classical surface (within $10\%$ along $[100]$ and $3\%$ along $[110]$) and to the zone boundary, while that of the mixed-sector UB31 state lies well inside it, at $0.71$--$0.79$ of the classical radius for $J_2/J_1\ge1$. The directions of largest intensity are X for $N_f=3/2$, and $\langle110\rangle$ and $\langle100\rangle$ (nearly degenerate) for the mixed-sector state. These can be compared with the classical thermal order-by-disorder selection for $J_2/J_1>2/3$, around the $(q,0,0)$ directions~\cite{Bergman-2007}. The modulation along the mean-field shells is weak, however, so the intensity is spread over a surface rather than concentrated at isolated wave vectors, as in the functional renormalization group study of Ref.~\cite{Buessen-2018}.

\section{Summary and outlook}
\label{sec:summary}
We have extended the projective symmetry group classification, which has underpinned the theory of spin-$1/2$ quantum spin liquids for twenty-five years, to fermionic partons of arbitrary spin $S$. Higher-spin candidate materials and synthetic magnets now have a systematic list of the symmetric spin liquids they can host. For $S\!>\!1/2$ the particle and hole representations are no longer equivalent, and the internal redundancy of the representation depends on whether the spin is integer or half-odd-integer. This leads to different invariant gauge groups. For integer spin they include $\mathrm{U}(1)\rtimes\mathbb{Z}_2$ and $\mathbb{Z}_2\times\mathbb{Z}_2$, and the latter has no analog in the spin-$1/2$ classification. Integer-spin quantum spin liquids can realize projective symmetries in ways that do not occur for spin $1/2$.

For the spin-$1$ diamond lattice with amplitudes on first- and second-neighbor bonds, the classification gives ten fully symmetric \textit{Ans\"atze}: two with IGG $\mathrm{U}(1)\rtimes\mathbb{Z}_2$, five with $\mathrm{U}(1)$, two with $\mathbb{Z}_2\times\mathbb{Z}_2$, and one with $\mathbb{Z}_2$. The hopping-only states have the link structure of the spin-$1/2$ classification but a different gauge structure, because the internal symmetry group is different. The paired states differ more: the spin-$1$ singlet pairing field is odd under site interchange, unlike the even singlet pairing of spin-$1/2$ Abrikosov fermions, and this restricts the allowed paired \textit{Ans\"atze} and changes the pattern of $\mathbb{Z}_2$ descendants.

We then studied the antiferromagnetic $J_1$--$J_2$ Heisenberg model with these \textit{Ans\"atze}, including the symmetry-allowed staggered onsite terms of UA31 and UB31. In the particle-hole-symmetric treatment ($N_f=3/2$) the second-neighbor amplitudes of all hopping states vanish identically at small $J_2/J_1$, and the lowest computed state is the $\mathrm{U}(1)\rtimes\mathbb{Z}_2$ state SA3 up to $J_2/J_1\simeq0.82$. The class-B state UB31 takes over beyond this point. It is first gapped, with nonzero second-neighbor and staggered onsite amplitudes and a small first-neighbor amplitude. For $J_2/J_1\gtrsim1.0$ the first-neighbor amplitude vanishes and the two FCC sublattices decouple, which leaves a gapless state of two independent FCC lattices with second-neighbor hopping, in which UB01 and UB31 are degenerate. This regime overlaps with the nonmagnetic regime found in earlier studies of the spin-$1$ diamond antiferromagnet.

The particle and hole representations, $N_f=1$ and $N_f=2$, are exact images of each other under the global $\mathbb{Z}_2$ gauge transformation and give identical phase diagrams. For UA31 and UB31 at fixed filling, the gauge parts of $S_4$ and $\bar C_3$ exchange the particle and hole representations between the sublattices. Their symmetric realizations therefore have the two sublattices in opposite constraint sectors. A uniform-sector realization is symmetric only if it is self-dual, as the decoupled class-B state is; otherwise, as for UA31 with $\chi^{}_2\neq0$, it breaks the sublattice-exchange symmetry (Appendix~\ref{app:selfdual}). UA01 and UB01 are symmetric only in the uniform sector, where the lowest state is UB01 up to $J_2/J_1\simeq1.37$ and UA01 above. The mixed sector lies lower at every coupling. A counting argument explains why the uniform-sector states are gapless: for integer spin, a symmetric $\mathrm{U}(1)$ state in the uniform sector can be gapped only if the number of sites per unit cell is a multiple of $2S+1$, whereas mixed-sector states on a bipartite lattice and paired states are not restricted. Among the symmetric realizations we computed, the lowest state is the paired \textit{Ansatz} ZA3110 at its self-consistent point $\chi^{}_1=0$, which is the mixed-sector realization of UA31, for $J_2/J_1\lesssim0.67$, and the mixed-sector realization of UB31 above. Both are gapped, and the $\mathbb{Z}_2$ \textit{Ansatz} adds no saddle point beyond the $\mathrm{U}(1)$ hopping states. In the particle-hole-averaged representation of Ref.~\cite{Liu-2010a}, where the internal $\mathbb{Z}_2$ leaves the constraint invariant and every saddle point is symmetric, the lowest states come in the same order as in the particle-hole-symmetric treatment; the window with $(\chi^{}_2,\lambda)\neq0$ widens, and its occupations move towards the mixed sector. The phase diagram, and properties of the candidate states such as whether nodal structures lie at the Fermi level, thus depend on how the spin-$1$ constraint is implemented.

This dependence is the main limitation of the saddle-point analysis. The particle-hole-symmetric treatment keeps the symmetry between the two representations, but it implements neither physical constraint exactly. The particle and hole representations impose the correct integer fillings at mean-field level, but they treat two inequivalent embeddings of the spin-$1$ Hilbert space separately. A natural next step is to study the classified \textit{Ans\"atze} with variational Monte Carlo, where the particle and hole constraints can be imposed exactly by projection. This would allow a direct comparison between the constraint representations and a stronger test of the candidate spin liquids identified here.

Restricting the classification to second-neighbor amplitudes also leaves some parent states without paired descendants. Within this restriction, the class-B $\mathrm{U}(1)$ states and the $\mathbb{Z}_2\times\mathbb{Z}_2$ states do not all have $\mathbb{Z}_2$ descendants. Extending the construction to third-neighbor bonds gives four further $\mathbb{Z}_2$ \textit{Ans\"atze}, listed in the last four rows of Table~\ref{table:ansatz} (Appendix~\ref{app:z2z2_psW_derivation}), but UB01 still has no $\mathbb{Z}_2$ descendant. The descendant structure of integer-spin PSGs thus appears to be more restrictive than for spin $1/2$, and longer-range amplitudes may be needed to reveal the full set of nearby paired spin liquids.

The classification also gives a systematic way into materials and models that have so far been treated case by case. Nothing in the construction is specific to the diamond lattice. It applies to any lattice and any spin, and the spin-$1/2$ classifications already available for the kagome~\cite{Lu-2011}, triangular~\cite{Lu-2016}, and pyrochlore~\cite{Liu-2021} lattices supply the link structure on which the integer-spin gauge structure can be built. Natural next targets are the spin-$1$ pyrochlore NaCaNi$_2$F$_7$, whose neutron spectra show a continuum of quantum fluctuations~\cite{Plumb-2019}, and the langbeinite K$_2$Ni$_2$(SO$_4$)$_3$, in which a field-induced spin-liquid regime on two coupled trillium lattices has been reported~\cite{Zivkovic-2021,Gonzalez-2024}. The counting rule of Sec.~\ref{sec:filling} already constrains both. The pyrochlore lattice has four sites per primitive cell and K$_2$Ni$_2$(SO$_4$)$_3$ has eight Ni$^{2+}$ ions per cubic cell; neither is a multiple of $2S+1=3$, so every time-reversal-symmetric spin-$1$ $\mathrm{U}(1)$ state in a uniform constraint sector has a spinon Fermi surface, and a gapped spin liquid in these materials must be paired or of mixed-sector type. For the spin-$3/2$ and spin-$5/2$ spinels CoAl$_2$O$_4$ and MnSc$_2$S$_4$~\cite{MacDougall-2011,Gao-2017}, the half-odd-integer hierarchy applies instead, and the same framework can address whether the spiral spin-liquid regime of the classical model~\cite{Bergman-2007} survives as a quantum spin liquid. For spin-$1$ Rydberg arrays~\cite{Liu-2024,Mogerle-2025}, the classified \textit{Ans\"atze} give candidate states to compare with quantum simulations. Together with projected wave functions and the structure factors computed from them, the spin-$S$ PSG can be used to identify fractionalized phases in higher-spin magnets generally.

\acknowledgments

R.S. was supported by the Princeton Quantum Initiative Fellowship. The work of Y.I. and R.S. was performed, in part, at the Aspen Center for Physics, which is supported by National Science Foundation Grant No.~PHY-2210452 and a grant from the Simons Foundation (1161654, Troyer). This research was supported in part by grant NSF PHY-2309135 to the Kavli Institute for Theoretical Physics (KITP). Y.I. acknowledges support from the ICTP through the Associates Programme, from the Simons Foundation through Grant No.~284558FY19, IIT Madras through the Institute of Eminence (IoE) program for establishing QuCenDiEM (Project No.~SP22231244CPETWOQCDHOC), and the International Centre for Theoretical Sciences (ICTS), Bengaluru  during a visit for participating in the Discussion Meeting - Fractionalized Quantum Matter (code: ICTS/DMFQM2025/07). Y.I. also acknowledges the use of the computing resources at HPCE, IIT Madras. The work in W\"urzburg is supported by the Deutsche Forschungsgemeinschaft (DFG, German Research Foundation) through Project-ID 258499086 - SFB 1170 and through the W\"urzburg-Dresden Cluster of Excellence on Complexity, Topology and Dynamics in Quantum Matter - ctd.qmat Project-ID 390858490 - EXC 2147.

\section*{Data Availability Statement} All codes and data underlying the results of this paper, including the scripts that reproduce every figure and the self-consistent mean-field parameters, will be deposited in a public repository (Zenodo) upon publication; until then they are available from the corresponding author.

\appendix

\section{Generic algebraic PSG conditions}
\label{app:gauge_conditions}

In this appendix, we collect the algebraic PSG constraints used in the diamond-lattice classification.  The diamond lattice is generated by the three translations $T_1,T_2,T_3$, the screw rotation $S_4$, and the sixfold rotoinversion $\bar C_3$.  We use the same convention as in the main text, where $T_i$ translates the lattice along the $i$th FCC primitive direction and $i+1$ is understood cyclically,
\begin{equation}
i+1\equiv {\rm mod}(i,3)+1 .
\end{equation}
The ordinary multiplication rules among the space-group generators are
\begin{subequations}
\label{eq:diamond_group_relations}

\begin{align}
T_iT_{i+1}T_i^{-1}T_{i+1}^{-1}
&= I,
\label{eq:group_translation_commutation}\\
T_i\bar C_3^{-1}T_{i+1}\bar C_3
&= I,
\label{eq:group_C3_translation}\\
\bar C_3^{\,6}
&= I .
\label{eq:group_C3_six}
\end{align}

\begin{align}
T_1^{-1}S_4T_2S_4^{-1}
&= I,
\label{eq:group_S4_T1}\\
T_2^{-1}S_4T_2T_3^{-1}S_4^{-1}
&= I,
\label{eq:group_S4_T2}\\
S_4T_1^{-1}T_2S_4^{-1}T_3^{-1}
&= I,
\label{eq:group_S4_T3}\\
S_4T_1S_4^{-1}T_1^{-1}T_3
&= I,
\label{eq:group_S4_T4}\\
S_4T_3^{-1}S_4^{-1}T_2^{-1}T_1
&= I,
\label{eq:group_S4_T5}\\
S_4^4T_1^{-1}T_2^{-1}T_3
&= I .
\label{eq:group_S4_four}
\end{align}

\begin{align}
(S_4\bar C_3)^2
&= I,
\label{eq:group_S4_C3_1}\\
(\bar C_3^{\,2}S_4^2)^3
&= I,
\label{eq:group_S4_C3_2}\\
(\bar C_3^{\,3}S_4)^4
&= I,
\label{eq:group_S4_C3_3}\\
(\bar C_3S_4^2)^6
&= I .
\label{eq:group_S4_C3_4}
\end{align}

\end{subequations}
Time reversal does not move lattice sites and so commutes
with all space-group generators.  We take
\begin{subequations}
\label{eq:time_reversal_group_relations}
\begin{align}
\mathcal T^2
&= I,
\label{eq:group_time_square}\\
\mathcal T\,\mathcal O\,\mathcal T^{-1}\mathcal O^{-1}
&= I,
\qquad
\mathcal O\in\{T_1,T_2,T_3,S_4,\bar C_3\}.
\label{eq:group_time_commutation}
\end{align}
\end{subequations}

In the PSG construction, each physical symmetry operation $\mathcal O$ is accompanied by a gauge transformation $W^{}_{\mathcal O}$.  A relation among physical symmetry operations is then required to hold only up to an element of the invariant gauge group.  Thus, if
\begin{equation}
\mathcal O_1\mathcal O_2\cdots \mathcal O_n=I
\end{equation}
is a relation in the physical symmetry group, its projective implementation obeys
\begin{equation}
(W_{\mathcal O_1}\mathcal O_1)
(W_{\mathcal O_2}\mathcal O_2)
\cdots
(W_{\mathcal O_n}\mathcal O_n)
\in {\rm IGG}.
\end{equation}
Applying this rule to Eqs.~\eqref{eq:diamond_group_relations}
and~\eqref{eq:time_reversal_group_relations} gives the following
algebraic PSG constraints:
\begin{subequations}
\label{eq:id_gauge_relation}

\begin{align}
&(W_{T_i}T_i)
(W_{T_{i+1}}T_{i+1})
(W_{T_i}T_i)^{-1}
(W_{T_{i+1}}T_{i+1})^{-1}
\in {\rm IGG},
\label{eq:gauge_translation_commutation}\\
&(W_{T_i}T_i)
(W_{\bar C_3}\bar C_3)^{-1}
(W_{T_{i+1}}T_{i+1})
(W_{\bar C_3}\bar C_3)
\in {\rm IGG},
\label{eq:gauge_C3_translation}\\
&(W_{\bar C_3}\bar C_3)^6
\in {\rm IGG}.
\label{eq:gauge_C3_six}
\end{align}

\begin{align}
&(W_{T_1}T_1)^{-1}
(W_{S_4}S_4)
(W_{T_2}T_2)
(W_{S_4}S_4)^{-1}
\in {\rm IGG},
\label{eq:gauge_S4_T1}\\
&(W_{T_2}T_2)^{-1}
(W_{S_4}S_4)
(W_{T_2}T_2)
(W_{T_3}T_3)^{-1}
(W_{S_4}S_4)^{-1}
\in {\rm IGG},
\label{eq:gauge_S4_T2}\\
&(W_{S_4}S_4)
(W_{T_1}T_1)^{-1}
(W_{T_2}T_2)
(W_{S_4}S_4)^{-1}
(W_{T_3}T_3)^{-1}
\in {\rm IGG},
\label{eq:gauge_S4_T3}\\
&(W_{S_4}S_4)
(W_{T_1}T_1)
(W_{S_4}S_4)^{-1}
(W_{T_1}T_1)^{-1}
(W_{T_3}T_3)
\in {\rm IGG},
\label{eq:gauge_S4_T4}\\
&(W_{S_4}S_4)
(W_{T_3}T_3)^{-1}
(W_{S_4}S_4)^{-1}
(W_{T_2}T_2)^{-1}
(W_{T_1}T_1)
\in {\rm IGG},
\label{eq:gauge_S4_T5}\\
&(W_{S_4}S_4)^4
(W_{T_1}T_1)^{-1}
(W_{T_2}T_2)^{-1}
(W_{T_3}T_3)
\in {\rm IGG}.
\label{eq:gauge_S4_four}
\end{align}

\begin{align}
&\left[(W_{S_4}S_4)
(W_{\bar C_3}\bar C_3)\right]^2
\in {\rm IGG},
\label{eq:gauge_S4_C3_1}\\
&\left[(W_{\bar C_3}\bar C_3)^2
(W_{S_4}S_4)^2\right]^3
\in {\rm IGG},
\label{eq:gauge_S4_C3_2}\\
&\left[(W_{\bar C_3}\bar C_3)^3
(W_{S_4}S_4)\right]^4
\in {\rm IGG},
\label{eq:gauge_S4_C3_3}\\
&\left[(W_{\bar C_3}\bar C_3)
(W_{S_4}S_4)^2\right]^6
\in {\rm IGG}.
\label{eq:gauge_S4_C3_4}
\end{align}

\begin{align}
&(W_{\mathcal T}\mathcal T)^2
\in {\rm IGG},
\label{eq:gauge_time_square}\\
&(W_{\mathcal T}\mathcal T)
(W^{}_{\mathcal O}\mathcal O)
(W_{\mathcal T}\mathcal T)^{-1}
(W^{}_{\mathcal O}\mathcal O)^{-1}
\in {\rm IGG},
\nonumber\\
&\hspace{3.0cm}
\mathcal O\in
\{T_1,T_2,T_3,S_4,\bar C_3\}.
\label{eq:gauge_time_commutation}
\end{align}

\end{subequations}
These equations are the starting point for the algebraic PSG classification.  The meaning of the right-hand side depends on the chosen IGG.  For instance, for a $\mathbb{Z}_2$ IGG, it is represented by $\pm\tau^0$, while for a $\mathrm{U}(1)$ IGG, it is represented by an element of the form $e^{i\phi\tau^3}$.  For integer-spin \textit{Ans\"atze}, the possible IGGs include the additional structures $\mathrm{U}(1)\rtimes\mathbb{Z}_2$ and $\mathbb{Z}_2\times\mathbb{Z}_2$, as discussed in the main text.

\section{$\mathrm{U}(1)\rtimes\mathbb{Z}_2$ and $\mathrm{U}(1)$ PSGs}
\label{app:u1z2_u1_psg}

We first discuss the PSG solutions whose canonical \textit{Ans\"atze} have IGG $\mathrm{U}(1)\rtimes\mathbb{Z}_2$ or $\mathrm{U}(1)$.  In the canonical gauge, the corresponding link matrices may be chosen as
\begin{equation}
\label{eq:canonical_u1z2_u1}
\begin{aligned}
{\rm IGG}=\mathrm{U}(1)\rtimes\mathbb{Z}_2:
\quad
& u_{ij}=i\chi_{ij}\tau^0,\\
{\rm IGG}=\mathrm{U}(1):
\quad
& u_{ij}=i\chi^0_{ij}\tau^0+\chi^3_{ij}\tau^3 .
\end{aligned}
\end{equation}
Here, $\tau^0$ is the $2\times2$ identity matrix in the internal two-column space, and $\chi^{}_{ij}$, $\chi^0_{ij}$, and $\chi^3_{ij}$ are real mean-field amplitudes in this canonical parametrization.

The gauge transformations associated with a symmetry operation $\mathcal O$ must preserve the appropriate canonical form in Eq.~\eqref{eq:canonical_u1z2_u1}.  They can then be written as
\begin{equation}
\label{eq:canonical_gauge_u1z2_u1}
\begin{aligned}
{\rm IGG}=\mathrm{U}(1)\rtimes\mathbb{Z}_2:\quad
&W^{}_{\mathcal O}(r,\mu)
=
\eta^{}_{\mathcal O}(r,\mu)\,
\mathcal{W}^{}_{\mathcal O}(\mu),\\
{\rm IGG}=\mathrm{U}(1):\quad
&W^{}_{\mathcal O}(r,\mu)
=
e^{i\phi_{\mathcal O}(r,\mu)\tau^3}
(\tau^1)^{w_{\mathcal O}} .
\end{aligned}
\end{equation}
where $r\!=\!(r_1,r_2,r_3)$ is the unit-cell coordinate, $\mu=0,1$ labels the two diamond sublattices, and $\eta^{}_{\mathcal O}(r,\mu)=\pm1$. In the first line, $\eta_{\mathcal O}(r,\mu)=\pm1$ is a
site-dependent sign and
\begin{equation}
\begin{aligned}
\mathcal{W}^{}_{\mathcal O}(\mu)
&\in \mathrm{U}(1)\rtimes\mathbb{Z}_2  \\
&=
\left\{
e^{i\phi\tau^3},
e^{i\phi\tau^3}\tau^1
\right\}.
\end{aligned}
\end{equation}
In the second line, $\phi_{\mathcal O}(r,\mu)$ is a real phase and
\begin{equation}
w^{}_{\mathcal O}=0,1 .
\end{equation}
The symmetry operations include
\begin{equation}
\mathcal O\in\{T_1,T_2,T_3,\bar C_3,S_4\}.
\end{equation}

The structures of these canonical gauges closely parallel the spin-$1/2$ diamond-lattice classification. The $\mathrm{U}(1)\rtimes\mathbb{Z}_2$ spin-$1$ \textit{Ans\"atze} play the role of the $\mathrm{SU}(2)$ \textit{Ans\"atze} in the spin-$1/2$ Abrikosov-fermion construction, with the important replacement
\begin{equation}
\begin{aligned}
\mathcal{W}^{}_{\mathcal O}(\mu)\in \mathrm{SU}(2)
\quad\longrightarrow\quad
\mathcal{W}^{}_{\mathcal O}(\mu)
&\in \mathrm{U}(1)\rtimes\mathbb{Z}_2 .
\end{aligned}
\end{equation}
Similarly, the $\mathrm{U}(1)$ PSG equations have the same algebraic structure as in the spin-$1/2$ case, except that the $\mathbb{Z}_2$-coset part of the gauge transformation is represented by $(\tau^1)^{w_{\mathcal O}}$ rather than $(i\tau^1)^{w_{\mathcal O}}$.  This difference reflects the integer-spin internal gauge structure, but it does not change the pattern of algebraic PSG solutions. Consequently, the $\mathrm{U}(1)\rtimes\mathbb{Z}_2$ and $\mathrm{U}(1)$ PSG solutions for spin-$1$ systems on the diamond lattice are obtained from the corresponding spin-$1/2$ $\mathrm{SU}(2)$ and $\mathrm{U}(1)$ solutions by this replacement of the allowed gauge group.

	\begin{table*}
	\caption{Gauge-inequivalent algebraic PSG solutions with $\mathbb{Z}_2$ IGG for the spin-$1$ diamond lattice. The entries list the allowed choices of translation signs, time-reversal signs, and sublattice gauge matrices after using the available gauge freedom. The notation follows the labeling scheme $\mathrm{ZA/B}\,\eta_{S_4\bar C_3}\eta_{\mathcal T} \mathcal{W}_{\mathcal T}(0)\mathcal{W}_{\bar C_3}(0)$ described in Appendix~\ref{app:z2_psW_derivation}. The enumeration gives $160$ algebraic $\mathbb{Z}_2$ PSGs.}
	\begin{ruledtabular}
		\begin{tabular}{ccccccccccc}
	$\eta_1$&$\eta_{S_4T_3}$&$\eta^{}_{\bar{C}_3T_1}$&$\eta_{\bar{C}_3T_2}$& $\mathcal{W}_\mathcal{T}(0)$& $\mathcal{W}_\mathcal{T}(1)$ &$\mathcal{W}_{S_4}(0)$ & $\mathcal{W}_{S_4}(1)$ &$\mathcal{W}_{\bar{C}_3}(0)$ &$\mathcal{W}_{\bar{C}_3}(1)$ &  Number of PSGs  \vspace*{0.05cm}\\%
			\hline
			$\pm1$&$\eta_1$&$+1$&$+1$& $\tau^0$& $\eta_\mathcal{T}\tau^0$ &$\tau^0$ & $\eta_{S_4}\tau^0$ &$\tau^0$ &$\eta_{\bar{C}_3}\tau^0$ & 16\\
			$\pm1$&$\eta_1$&$+1$&$+1$& $\tau^0$& $\eta_\mathcal{T}\tau^0$ &$\tau^0$ & $\eta_{S_4}\tau^0$ &$i\tau^3$ &$\eta_{\bar{C}_3}i\tau^3$ & 16\\
			$\pm1$&$\eta_1$&$+1$&$+1$& $\tau^0$& $\eta_\mathcal{T}\tau^0$ &$\tau^0$ & $\eta_{S_4}\tau^0$ &$\tau^1$ &$\eta_{\bar{C}_3}\tau^1$ & 16\\
			$\pm1$&$\eta_1$&$+1$&$+1$& $i\tau^3$& $\eta_\mathcal{T}i\tau^3$ &$\tau^0$ & $\eta_{S_4}\tau^0$ &$\tau^0$ &$\eta_{\bar{C}_3}\tau^0$ & 16\\
			$\pm1$&$\eta_1$&$+1$&$+1$& $i\tau^3$& $\eta_\mathcal{T}i\tau^3$ &$\tau^0$ & $\eta_{S_4}\tau^0$ &$i\tau^3$ &$\eta_{\bar{C}_3}i\tau^3$ & 16\\
			$\pm1$&$\eta_1$&$+1$&$+1$& $i\tau^3$& $\eta_\mathcal{T}i\tau^3$ &$\tau^0$ & $\eta_{S_4}\tau^0$ &$\tau^1$ &$\eta_{\bar{C}_3}\tau^1$ & 16\\
			$\pm1$&$\eta_1$&$+1$&$+1$& $\tau^1$& $\eta_\mathcal{T}\tau^1$ &$\tau^0$ & $\eta_{S_4}\tau^0$ &$\tau^0$ &$\eta_{\bar{C}_3}\tau^0$ & 16\\
			$\pm1$&$\eta_1$&$+1$&$+1$& $\tau^1$& $\eta_\mathcal{T}\tau^1$ &$\tau^0$ & $\eta_{S_4}\tau^0$ &$i\tau^3$ &$\eta_{\bar{C}_3}i\tau^3$ & 16\\
			$\pm1$&$\eta_1$&$+1$&$+1$& $\tau^1$& $\eta_\mathcal{T}\tau^1$ &$\tau^0$ & $\eta_{S_4}\tau^0$ &$\tau^1$ &$\eta_{\bar{C}_3}\tau^1$ & 16\\
			$\pm1$&$\eta_1$&$+1$&$+1$& $\tau^2$& $\eta_\mathcal{T}\tau^2$ &$\tau^0$ & $\eta_{S_4}\tau^0$ &$\tau^1$ &$\eta_{\bar{C}_3}\tau^1$ & 16\\
		\end{tabular}
	\end{ruledtabular}
	\label{table:z2_psg}
\end{table*}

\section{$\mathbb{Z}_2$ PSG}
\label{app:z2_psW_derivation}

In this appendix, we summarize the algebraic PSG solutions with IGG $\mathbb{Z}_2$.  These arise when the continuous part of the integer-spin $\mathrm{U}(1)\rtimes\mathbb{Z}_2$ gauge structure is broken completely down to its central $\mathbb{Z}_2$ subgroup. In this case, the link matrix must contain pairing components in addition to hopping components. In a canonical gauge, we may write
\begin{equation}
\label{eq:canonical_z2}
u_{ij}
=
i\chi^0_{ij}\tau^0
+\chi^3_{ij}\tau^3
+i\Delta^1_{ij}\tau^1
+i\Delta^2_{ij}\tau^2;
\end{equation}
the coefficients are real in this parametrization.  The terms proportional to $\tau^0$ and $\tau^3$ describe hopping channels, while those proportional to $\tau^1$ and $\tau^2$ describe singlet pairing channels.

For a $\mathbb{Z}_2$ IGG, the gauge transformations associated with a space-group or time-reversal operation $\mathcal O$ can be written in the form
\begin{equation}
\label{eq:canonical_gauge_z2}
W^{}_{\mathcal O}(r,\mu)
=
\eta^{}_{\mathcal O}(r,\mu)\,
\mathcal{W}^{}_{\mathcal O}(\mu).
\end{equation}
Here, $r\!=\!(r_1,r_2,r_3)$ denotes the unit-cell coordinate, $\mu=0,1$ labels the two diamond sublattices, and $\eta_{\mathcal O}(r,\mu)=\pm1$ is a site-dependent $\mathbb{Z}_2$ sign. The matrix $\mathcal{W}_{\mathcal O}(\mu)$ is sublattice dependent and belongs to the integer-spin internal gauge group,
\begin{equation}
\mathcal{W}^{}_{\mathcal O}(\mu)\in \mathrm{U}(1)\rtimes\mathbb{Z}_2
=
\left\{
e^{i\phi\tau^3},
e^{i\phi\tau^3}\tau^1
\right\}.
\end{equation}
This is the main difference from the spin-$1/2$ Abrikosov-fermion classification, where the corresponding sublattice matrices take values in $\mathrm{SU}(2)$.  Apart from this replacement of the allowed internal gauge group, the algebraic solution follows the same structure as in the spin-$1/2$ diamond-lattice PSG.

We solve the algebraic constraints listed in Appendix~\ref{app:gauge_conditions}.  After using gauge freedom to simplify the translational part, the $\mathbb{Z}_2$ PSG solutions may be written as
\begin{subequations}
\label{eq:z2_gauge_sol}

\begin{align}
W_{T_1}(r,\mu)
&=
\tau^0,
\label{eq:z2_GT1}\\
W_{T_2}(r,\mu)
&=
\eta_1^{\,r_1}\tau^0,
\label{eq:z2_GT2}\\
W_{T_3}(r,\mu)
&=
\eta_1^{\,r_1+r_2}\tau^0,
\label{eq:z2_GT3}\\
W_{\mathcal T}(r,\mu)
&=
\mathcal{W}_{\mathcal T}(\mu).
\label{eq:z2_GT}
\end{align}

\begin{align}
W_{S_4}(r,\mu)
&=
\eta_{S_4T_3}^{\,r_3}\,
\eta_1^{\,\Omega_{S_4}(r,\mu)}
\mathcal{W}_{S_4}(\mu),
\label{eq:z2_GS4}\\
\Omega_{S_4}(r,\mu)
&=
r_2\left(\frac{r_2-1}{2}+r_3-\bar\mu\right)
-
r_3\left(\frac{r_3-1}{2}+r_1\right),
\label{eq:z2_GS4_exponent}\\
{W}_{\bar C_3}(r,\mu)
&=
\eta_{\bar C_3T_1}^{\,r_2}\,
\eta_{\bar C_3T_2}^{\,r_3}\,
\eta_1^{\,r_1(r_2+r_3)}
\mathcal{W}_{\bar C_3}(\mu).
\label{eq:z2_GC3}
\end{align}

\end{subequations}
Here, $\bar\mu=1-\mu$, and all $\eta$ variables take values $\pm1$.  The remaining sublattice matrices
\begin{equation}
\mathcal{W}_{\mathcal T}(\mu),\qquad
\mathcal{W}_{S_4}(\mu),\qquad
\mathcal{W}_{\bar C_3}(\mu)
\end{equation}
are fixed by the point-group and time-reversal algebraic constraints, in particular those involving $\bar C_3^6$, $S_4^4$, the mixed $S_4$--$\bar C_3$ relations, and $\mathcal T^2$.

Carrying out this enumeration gives a total of $160$ algebraic $\mathbb{Z}_2$ PSGs.  We label them using the convention
\begin{equation}
\label{eq:z2_label}
\mathrm{ZA/B},
\eta_{S_4\bar C_3},
\eta_{\mathcal T},
\mathcal{W}_{\mathcal T}(0),
\mathcal{W}_{\bar C_3}(0).
\end{equation}
The initial letter $A$ or $B$ distinguishes the two translational classes.  The symbol $\eta_{S_4\bar C_3}$ encodes the pair of signs $(\eta_{S_4},\eta_{\bar C_3})$ according to
\begin{equation}
\label{eq:z2_eta_label}
\eta_{S_4\bar C_3}
=
\begin{cases}
0, & (\eta_{S_4},\eta_{\bar C_3})=(+1,+1),\\
1, & (\eta_{S_4},\eta_{\bar C_3})=(+1,-1),\\
2, & (\eta_{S_4},\eta_{\bar C_3})=(-1,+1),\\
3, & (\eta_{S_4},\eta_{\bar C_3})=(-1,-1).
\end{cases}
\end{equation}
Similarly, we write $0$ and $1$ for the two possible signs of $\eta_{\mathcal T}$, with $0$ denoting $+1$ and $1$ denoting $-1$.  If
\begin{equation}
\mathcal{W}_{\bar C_3}(0)\propto \tau^\alpha ,
\end{equation}
we write the index $\alpha$ in the label; the same convention is used for $\mathcal{W}_{\mathcal T}(0)$.  The complete list of the $160$ algebraic $\mathbb{Z}_2$ PSGs is given in Table~\ref{table:z2_psg}.

	\begin{table*}
	\caption{Gauge-inequivalent algebraic PSG solutions with $\mathbb{Z}_2\times\mathbb{Z}_2$ IGG. The binary variables $w_{S_4}$, $w_{\bar C_3}$, and $w_{\mathcal T}$ specify whether the corresponding symmetry operation is accompanied by the additional gauge transformation $i\tau^3$ (see text). An entry $a/b$ denotes two allowed choices, one for each of two distinct PSGs; entries $0,1$ denote both values of a binary parameter. The remaining columns specify the allowed signs, $\epsilon$ parameters, and sublattice matrices $\mathcal{W}_{\mathcal O}(\mu)$ after fixing the residual gauge freedom. The full enumeration gives $512$ algebraic PSGs of $\mathbb{Z}_2\times\mathbb{Z}_2$ type.}
	\begin{ruledtabular}
		\begin{tabular}{cccccccccccccc}
	$\eta_1$&$w_{S_4}$&$w_{\bar{C}_3}$&$w_{\mathcal{T}}$&$\epsilon_{T_1}$&$\epsilon_{S_4T_2}$&$\epsilon_{\bar{C}_3T_2}$& $\mathcal{W}_\mathcal{T}(0)$& $\mathcal{W}_\mathcal{T}(1)$ &$W_{S_4}(0)$ & $\mathcal{W}_{S_4}(1)$ &$\mathcal{W}_{\bar{C}_3}(0)$ &$\mathcal{W}_{\bar{C}_3}(1)$ &  Number of PSGs \vspace*{0.05cm}\\%
			\hline
			$\pm1$&$0$&$0$&$0,1$&$0,1$&$0$&$0$& $\tau^0$& $-\tau^0/\tau^1$ &$\tau^0$ & $\eta_{S_4}\tau^0$ &$\tau^0$ &$\eta_{\bar{C}_3}\tau^0$ & 64\\
			$\pm1$&$0$&$0$&$0,1$&$0,1$&$0$&$0$& $\tau^0$& $-\tau^0/\tau^1$ &$\tau^0$ & $\eta_{S_4}\tau^0$ &$\tau^1$ &$\eta_{\bar{C}_3}\tau^1$ & 64\\
			$\pm1$&$0$&$0$&$0$&$0,1$&$0$&$0$& $\tau^0$& $-\tau^0/\tau^1$ &$\tau^0$ & $\eta_{S_4}\tau^0$ &$\tau^0$ &$\eta_{\bar{C}_3}\tau^1$ & 32\\
			$\pm1$&$0$&$0$&$0$&$0,1$&$0$&$0$& $\tau^0$& $-\tau^0/\tau^1$ &$\tau^0$ & $\eta_{S_4}\tau^1$ &$\tau^0$ &$\eta_{\bar{C}_3}\tau^0$ & 32\\
			$\pm1$&$0$&$0$&$0$&$0,1$&$0$&$0$& $\tau^0$& $-\tau^0/\tau^1$ &$\tau^0$ & $\eta_{S_4}\tau^1$ &$\tau^1$ &$\eta_{\bar{C}_3}\tau^1$ & 32\\
			$\pm1$&$0$&$0$&$0$&$0,1$&$0$&$0$& $\tau^0$& $-\tau^0/\tau^1$ &$\tau^0$ & $\eta_{S_4}\tau^1$ &$\tau^0$ &$\eta_{\bar{C}_3}\tau^1$ & 32\\
			\hline
			$\pm1$&$1$&$0$&$0$&$0$&$0,1$&$0,1$& $\tau^0$& $-\tau^0$ &$\tau^0$ & $\eta_{S_4}\tau^0$ &$\tau^{0}/\tau^{1}$ &$\eta_{\bar{C}_3}\tau^{0}/\tau^{1}$ & 64\\
			$\pm1$&$1$&$0$&$1$&$0$&$0,1$&$0$& $\tau^0$& $\eta_\mathcal{T}\tau^0$ &$\tau^0$ & $\eta_{S_4}\tau^0$ &$\tau^{0}/\tau^{1}$ &$\eta_{\bar{C}_3}\tau^{0}/\tau^{1}$ & 64\\
			\hline
			$\pm1$&$0$&$1$&$0$&$0$&$0,1$&$0$& $\tau^0$& $-\tau^0$ &$\tau^0$ & $\eta_{S_4}\tau^0$ &$\tau^{0}/\tau^{1}$ &$\eta_{\bar{C}_3}\tau^{0}/\tau^{1}$ & 32\\
			$\pm1$&$0$&$1$&$1$&$0$&$0$&$0$& $\tau^0$& $\eta_\mathcal{T}\tau^0$ &$\tau^0$ & $\eta_{S_4}\tau^0$ &$\tau^{0}/\tau^{1}$ &$\eta_{\bar{C}_3}\tau^{0}/\tau^{1}$ & 32\\
			\hline
			$\pm1$&$1$&$1$&$0$&$0$&$0,1$&$0$& $\tau^0$& $-\tau^0$ &$\tau^0$ & $\eta_{S_4}\tau^0$ &$\tau^{0}/\tau^{1}$ &$\eta_{\bar{C}_3}\tau^{0}/\tau^{1}$ & 32\\
			$\pm1$&$1$&$1$&$1$&$0$&$0$&$0$& $\tau^0$& $\eta_\mathcal{T}\tau^0$ &$\tau^0$ & $\eta_{S_4}\tau^0$ &$\tau^{0}/\tau^{1}$ &$\eta_{\bar{C}_3}\tau^{0}/\tau^{1}$ & 32\\
		\end{tabular}
	\end{ruledtabular}
	\label{table:z2z2_psg}
\end{table*}

\section{$\mathbb{Z}_2\times\mathbb{Z}_2$ PSG}
\label{app:z2z2_psW_derivation}

Next, we describe the algebraic PSG solutions with IGG $\mathbb{Z}_2\times\mathbb{Z}_2$.  This class is special to the integer-spin fermionic construction.  In the canonical gauge, the IGG is the Klein four-group generated by the two commuting matrices $-\tau^0$ and $\tau^1$,
\begin{equation}
{\rm IGG}
=
\left\{
\tau^0,\,-\tau^0,\,\tau^1,\,-\tau^1
\right\}.
\end{equation}
A link matrix invariant under this IGG can be chosen in the form
\begin{equation}
\label{eq:canonical_z2z2}
u_{ij}
=
i\left(\chi_{ij}\tau^0+\Delta_{ij}\tau^1\right),
\end{equation}
where $\chi_{ij}$ and $\Delta_{ij}$ are real mean-field amplitudes.  The coexistence of the $\tau^0$ and $\tau^1$ components breaks the continuous $\mathrm{U}(1)$ part of the integer-spin $\mathrm{U}(1)\rtimes\mathbb{Z}_2$ structure down to a discrete subgroup, while retaining the additional $\tau^1$ element.  This gives the IGG $\mathbb{Z}_2\times\mathbb{Z}_2$.

The gauge transformations that preserve the canonical form in Eq.~\eqref{eq:canonical_z2z2} can be written as
\begin{equation}
\label{eq:canonical_z2z2_gauge_structure}
\begin{aligned}
W^{}_{\mathcal O}(r,\mu)
&=
\eta^{}_{\mathcal O}(r,\mu)\,
(\tau^1)^{f_{\mathcal O}(r,\mu)}
(i\tau^3)^{w_{\mathcal O}} \\
&\equiv
W_{\mathcal O}(r,\mu)
(i\tau^3)^{w_{\mathcal O}} ,
\end{aligned}
\end{equation}
  The exponent $f_{\mathcal O}(r,\mu)$ takes values in $\{0,1\}$ and is understood modulo two.  The binary variable
\begin{equation}
w_{\mathcal O}=0,1
\end{equation}
keeps track of whether the symmetry operation $\mathcal O$ is accompanied by the additional gauge operation $i\tau^3=e^{i\pi\tau^3/2}$. This element lies in the connected $\mathrm{U}(1)$ component of the internal group but outside the IGG.  We use
\begin{equation}
\mathcal O\in\{T_1,T_2,T_3,S_4,\bar C_3,\mathcal T\}.
\end{equation}

We first solve the translational part.  For $\mathcal O\!=\!T_i$, Eq. \eqref{eq:canonical_z2z2_gauge_structure} yields
\begin{equation}
\begin{aligned}
W_{T_i}(r,\mu)
&=
W_{T_i}(r,\mu)
(i\tau^3)^{w_{T_i}} .
\end{aligned}
\end{equation}
The possible values of $(w_{T_1},w_{T_2},w_{T_3})$ can be reduced, using the permutation symmetry among translations, to four cases,
\begin{equation}
(0,0,0),\qquad
(1,0,0),\qquad
(1,1,0),\qquad
(1,1,1).
\end{equation}
The last three choices are incompatible with the translation--point-group algebraic PSG constraints in Appendix~\ref{app:gauge_conditions}. Thus, only
\begin{equation}
w_{T_1}=w_{T_2}=w_{T_3}=0
\end{equation}
is allowed.  We can then use gauge freedom to impose
\begin{equation}
\begin{aligned}
W_{T_1}(r,\mu)&=\tau^0,\\
W_{T_2}(0,r_2,r_3,\mu)&=\tau^0,\\
W_{T_3}(0,0,r_3,\mu)&=\tau^0 .
\end{aligned}
\end{equation}
The translation commutation relation then gives
\begin{subequations}
\label{eq:z2z2_translation_gauges}
\begin{align}
W_{T_1}(r,\mu)
&=
\tau^0,
\label{eq:z2z2_GT1}\\
W_{T_2}(r,\mu)
&=
\eta_1^{\,r_1}
(\tau^1)^{\epsilon_{T_1}r_1},
\label{eq:z2z2_GT2}\\
W_{T_3}(r,\mu)
&=
\eta_1^{\,r_1+r_2}
(\tau^1)^{\epsilon_{T_1}(r_1+r_2)} .
\label{eq:z2z2_GT3}
\end{align}
\end{subequations}
Here, $\eta_1=\pm1$ and $\epsilon_{T_1}=0,1$.  In writing Eq.~\eqref{eq:z2z2_translation_gauges}, we have already used the consistency conditions
\begin{equation}
\eta_1=\eta_2=\eta_3,
\qquad
\epsilon_{T_1}=\epsilon_{T_2}=\epsilon_{T_3},
\end{equation}
which follow from the compatibility of translations with $\bar C_3$.

We next determine the $\bar C_3$ gauge. Define
\begin{equation}
\begin{aligned}
W_{\bar C_3}(r,\mu)
&=
W_{\bar C_3}(r,\mu)
(i\tau^3)^{w_{\bar C_3}} ,
\end{aligned}
\end{equation}
and introduce the discrete derivative
\begin{equation}
\begin{aligned}
f_i[W_{\bar C_3}(r,\mu)]
&=
W_{\bar C_3}(r,\mu)
W_{\bar C_3}^{-1}[T_i^{-1}(r,\mu)] .
\end{aligned}
\end{equation}
Substituting the translation solution into the $\bar C_3$--translation
relations gives
\begin{subequations}
\label{eq:z2z2_C3_fi}
\begin{align}
f_1[W_{\bar C_3}(r,\mu)]
&=
\eta_{\bar C_3T_3}
\eta_1^{\,r_2+r_3}
(\tau^1)^{
\epsilon_{\bar C_3T_3}
+\epsilon_{T_1}(r_2+r_3)
}
\nonumber\\
&\qquad\times
(-1)^{w_{\bar C_3}\epsilon_{T_1}(r_2+r_3)},
\label{eq:z2z2_C3_f1}\\
f_2[W_{\bar C_3}(r,\mu)]
&=
\eta_{\bar C_3T_1}
\eta_1^{\,r_1}
(\tau^1)^{
\epsilon_{\bar C_3T_1}
+\epsilon_{T_1}r_1
},
\label{eq:z2z2_C3_f2}\\
f_3[W_{\bar C_3}(r,\mu)]
&=
\eta_{\bar C_3T_2}
\eta_1^{\,r_1+r_2}
(\tau^1)^{
\epsilon_{\bar C_3T_2}
+\epsilon_{T_1}(r_1+r_2)
}
\nonumber\\
&\qquad\times
(-1)^{w_{\bar C_3}\epsilon_{T_1}r_2}.
\label{eq:z2z2_C3_f3}
\end{align}
\end{subequations}
The discrete derivatives must satisfy the integrability condition
\begin{equation}
\label{eq:z2z2_consistency}
\begin{aligned}
&f_i[W_{\bar C_3}(r,\mu)]\,
f_{i+1}[W_{\bar C_3}(T_i^{-1}r,\mu)]
\\
&\qquad =
f_{i+1}[W_{\bar C_3}(r,\mu)]\,
f_i[W_{\bar C_3}(T_{i+1}^{-1}r,\mu)] .
\end{aligned}
\end{equation}
This condition imposes
\begin{equation}
(-1)^{w_{\bar C_3}\epsilon_{T_1}}=1 .
\end{equation}
The resulting solution for the $\bar C_3$ gauge is
\begin{align}
W_{\bar C_3}(r,\mu)
&=
\eta_{\bar C_3T_3}^{\,r_1}
\eta_{\bar C_3T_1}^{\,r_2}
\eta_{\bar C_3T_2}^{\,r_3}
\eta_1^{\,r_1(r_2+r_3)}
\nonumber\\
&\quad\times
(\tau^1)^{
\epsilon_{\bar C_3T_3}r_1
+\epsilon_{\bar C_3T_1}r_2
+\epsilon_{\bar C_3T_2}r_3
+\epsilon_{T_1}r_1(r_2+r_3)
}
\nonumber\\
&\quad\times
\mathcal{W}_{\bar C_3}(\mu),
\label{eq:z2z2_GC3}
\end{align}
where
\begin{equation}
\mathcal{W}_{\bar C_3}(\mu)
=
W_{\bar C_3}(0,\mu)
(i\tau^3)^{w_{\bar C_3}} .
\end{equation}
The relation $(\bar C_3)^6=I$, imposed projectively, further constrains the sublattice matrices.  For $w_{\bar C_3}=0$, one obtains
\begin{equation}
\label{eq:z2z2_C3_six_w0}
\left[
\mathcal{W}_{\bar C_3}(\mu)
\mathcal{W}_{\bar C_3}(\bar\mu)
\right]^3
=
\eta_{\bar C_3}
(\tau^1)^{\epsilon_{\bar C_3}},
\end{equation}
whereas for $w_{\bar C_3}=1$, one obtains
\begin{equation}
\label{eq:z2z2_C3_six_w1}
\left[
\mathcal{W}_{\bar C_3}(\mu)
\mathcal{W}_{\bar C_3}(\bar\mu)
\right]^3
=
\eta_{\bar C_3}
(\tau^1)^{\epsilon_{\bar C_3}},
\end{equation}
together with
\begin{equation}
\epsilon_{\bar C_3T_1}
+
\epsilon_{\bar C_3T_2}
+
\epsilon_{\bar C_3T_3}
=
0
\quad
(\mathrm{mod}\;2).
\end{equation}
Here, $\bar\mu=1-\mu$, and $\eta_{\bar C_3}=\pm1$, $\epsilon_{\bar C_3}=0,1$ are IGG-valued constants.

\begin{table*}[tp]
\begin{ruledtabular}
\begin{tabular}{ccccc}
\multirow{3}{*}{Label} & 1NN & 2NN & 3NN & Onsite\\
& \{$u_{OA},u_{OB},u_{OC},u_{OD}$\} & \{$u^0_{O1},u^0_{O2},u^0_{O3},u^0_{O4},u^0_{O5},u^0_{O6},$ & \{$u_{Oa},u_{Ob},u_{Oc},u_{Od},u_{Oe},u_{Of},$ & $\{\lambda(0),\lambda(1)\}$  \\
& & \,\,\,\,$u^1_{O1},u^1_{O2},u^1_{O3},u^1_{O4},u^1_{O5},u^1_{O6}\}$ & \,\,\,\,$u_{Og},u_{Oh},u_{Oi},u_{Oj},u_{Ok},u_{Ol}\}$ & \\
\hline
$\mathrm{ZZA30}$& $\{u_1=\chi^{}_1\tau^3+i\Delta^{}_1\tau^1\}$ & $\{u_2=0\}$ & $\{u_3=\chi^{}_3\tau^3+i\Delta_3\tau^1\}$ & $\{0\}$ \\
$\mathrm{ZZB10}$& $\{u_1=\chi^{}_1\tau^3+i\Delta^{}_1\tau^1\}$ & $\{u_2=0\}$ & $\{u_3=0\}$ & $\{0\}$ \\

$\mathrm{ZZB00}$& $\{u_1=\chi^{}_1\tau^3\}$ & $\{u_2=0\}$ & $\{u_3=i\Delta_3\tau^1,u_3,-u_3,u_3,u_3,-u_3,$ & $\{0\}$ \\
&& & $-u_3,-u_3,-u_3,-u_3,-u_3,-u_3\}$ & \\

$\mathrm{ZA3110}$& $\{u_1=\chi^{}_1\tau^3+i\Delta^{}_1\tau^1\}$ & $\{u_2=\chi^{}_2\tau^3\}$ & $\{u_3=\chi^{}_3\tau^3+i\Delta_3\tau^1\}$ & $\{\lambda\}$ \\

$\mathrm{ZA3100}$& $\{u_1=\chi^{}_1\tau^3+i\Delta^{}_1\tau^1\}$ & $\{u_2=0\}$ & $\{u_3=\chi^{}_3\tau^3+i\Delta_3\tau^1+i\Delta'_3\tau^2\}$ & $\{0\}$ \\

$\mathrm{ZB0113}$& $\{u_1=\chi^{}_1\tau^3\}$ & $\{u_2=\chi^{}_2\tau^3,-u_2,u_2,u_2,u_2,-u_2,$ & $\{u_3=i\Delta_3\tau^1,u_3,-u_3,u_3,u_3,-u_3,$ & $\{\lambda,-\lambda\}$ \\
&&$u'_2=-u_2,u'_2,u'_2,-u'_2,u'_2,u'_2\}$ & $-u_3,-u_3,-u_3,-u_3,-u_3,-u_3\}$ & \\

$\mathrm{ZB0101}$& $\{u_1=\chi^{}_1\tau^3\}$ & $\{u_2=0\}$ & $\{u_3=i\chi^{}_3\tau^0+i\Delta_3\tau^1,u_3,-u_3,u_3,u_3,-u_3,$ & $\{0\}$ \\
& & & $-u_3,-u_3,-u_3,-u_3,-u_3,-u_3\}$ & \\

$\mathrm{ZB1101}$& $\{u_1=\chi^{}_1\tau^3+i\Delta^{}_1\tau^1\}$ & $\{u_2=0\}$ & $\{u_3=i\Delta_3\tau^2,u_3,-u_3,u_3,u_3,-u_3,$ & $\{0\}$ \\
& & & $-u_3,-u_3,-u_3,-u_3,-u_3,-u_3\}$ & \\
\end{tabular}
\end{ruledtabular}
\caption{Short-range $\mathbb{Z}_2\times\mathbb{Z}_2$ \textit{Ans\"atze} and nearby $\mathbb{Z}_2$ descendants on the diamond lattice with amplitudes retained up to third-neighbor bonds. The first-, second-, and third-neighbor columns list the symmetry-related link matrices generated from the reference site $O$ using the bond convention of Table~\ref{table:ansatz-j1-j2}. The table shows that including third-neighbor amplitudes produces additional paired $\mathbb{Z}_2$ descendants, while some parent states remain without short-range descendants.}
\label{table:ansatz}
\end{table*}

We now turn to the screw symmetry $S_4$.  Write
\begin{equation}
\begin{aligned}
W_{S_4}(r,\mu)
&=
W_{S_4}(r,\mu)
(i\tau^3)^{w_{S_4}} .
\end{aligned}
\end{equation}
The relations between $S_4$ and translations give
\begin{subequations}
\label{eq:z2z2_S4_fi}
\begin{align}
f_1[W_{S_4}(r,\mu)]
&=
\eta_{S_4T_1}
\eta_1^{\,r_3}
(-1)^{w_{S_4}\epsilon_{T_1}r_3}
(\tau^1)^{\epsilon_{S_4T_1}+\epsilon_{T_1}r_3},
\label{eq:z2z2_S4_f1}\\
f_2[W_{S_4}(r,\mu)]
&=
\eta_{S_4T_2}
\eta_1^{\,r_2+r_3-1-\bar\mu}
(-1)^{w_{S_4}\epsilon_{T_1}(r_2+r_3-1-\bar\mu)}
\nonumber\\
&\qquad\times
(\tau^1)^{
\epsilon_{S_4T_2}
+\epsilon_{T_1}(r_2+r_3-1-\bar\mu)
},
\label{eq:z2z2_S4_f2}\\
f_3[W_{S_4}(r,\mu)]
&=
\eta_{S_4T_3}
\eta_1^{\,r_1+r_2+r_3-1}
(-1)^{w_{S_4}\epsilon_{T_1}(r_3-1)}
\nonumber\\
&\qquad\times
(\tau^1)^{
\epsilon_{S_4T_3}
+\epsilon_{T_1}(r_1+r_2+r_3-1)
}.
\label{eq:z2z2_S4_f3}
\end{align}
\end{subequations}
The corresponding integrability conditions do not impose further restrictions.  A solution is
\begin{align}
W_{S_4}(r,\mu)
&=
\eta_{S_4T_1}^{\,r_1}
\eta_{S_4T_2}^{\,r_2}
\eta_{S_4T_3}^{\,r_3}
(\tau^1)^{
\epsilon_{S_4T_2}r_1
+\epsilon_{S_4T_1}r_2
+\epsilon_{S_4T_3}r_3
}
\nonumber\\
&\quad\times
\left[
\eta_1(\tau^1)^{\epsilon_{T_1}}
\right]^{\Omega_{S_4}(r,\mu)}
\mathcal{W}_{S_4}(\mu),
\label{eq:z2z2_GS4}\\
\Omega_{S_4}(r,\mu)
&=
r_3(r_1+r_2)
+
r_2\left(\frac{r_2-1}{2}-\bar\mu\right)
+
\frac{r_3(r_3-1)}{2}
\label{eq:z2z2_GS4_exponent}
\end{align}
with
\begin{equation}
\mathcal{W}_{S_4}(\mu)
=
W_{S_4}(0,\mu)
(i\tau^3)^{w_{S_4}} .
\end{equation}
The projective form of the $S_4^4$ relation gives additional
constraints. For $w_{S_4}=0$,
\begin{subequations}
\label{eq:z2z2_S4_four_w0}
\begin{align}
\eta_{S_4T_1}
\eta_{S_4T_2}
\eta_{S_4T_3}
&=
\eta_1,
\label{eq:z2z2_S4_four_w0_eta}\\
\epsilon_{S_4T_1}
+\epsilon_{S_4T_2}
+\epsilon_{S_4T_3}
&=
\epsilon_{T_1}
\quad
(\mathrm{mod}\;2).
\label{eq:z2z2_S4_four_w0_eps}
\end{align}
\end{subequations}

For $w_{S_4}=1$, the corresponding conditions become
\begin{subequations}
\label{eq:z2z2_S4_four_w1}
\begin{align}
\eta_{S_4T_1}
\eta_{S_4T_2}
\eta_{S_4T_3}
&=
\eta_1(-1)^{\epsilon_{S_4T_3}},
\label{eq:z2z2_S4_four_w1_eta}\\
\epsilon_{S_4T_1}
+\epsilon_{S_4T_2}
+\epsilon_{S_4T_3}
&=
\epsilon_{T_1}
=
0
\quad
(\mathrm{mod}\;2).
\label{eq:z2z2_S4_four_w1_eps}
\end{align}
\end{subequations}

The remaining point-group relations involving $S_4$ and $\bar C_3$, together with time reversal, further reduce the allowed choices of signs and sublattice matrices.  After using the residual $\mathrm{U}(1)\rtimes\mathbb{Z}_2$ gauge freedom, the gauge-inequivalent PSGs can be summarized compactly as follows.  Define
\begin{equation}
\bar w_{\mathcal O}=1-w_{\mathcal O},
\end{equation}
and
\begin{equation}
\label{eq:z2z2_fs_fc}
\begin{aligned}
f_s(r,\mu)
&=
r_2\left(\frac{r_2-1}{2}+r_3-\bar\mu\right)\\
&\quad
-
r_3\left(\frac{r_3-1}{2}+r_1+1\right),\\
f_c(r)
&=
r_1(r_2+r_3).
\end{aligned}
\end{equation}

For $w_{\mathcal T}=0$, the final gauge representatives may be written as
\begin{subequations}
\label{eq:z2z2_gauge_sol_wt0}
\begin{align}
W_{T_1}(r,\mu)
&=
\tau^0,
\label{eq:z2z2_wt0_GT1}\\
W_{T_2}(r,\mu)
&=
\eta_1^{\,r_1}
(\tau^1)^{\epsilon_{T_1}r_1},
\label{eq:z2z2_wt0_GT2}\\
W_{T_3}(r,\mu)
&=
\eta_1^{\,r_1+r_2}
(\tau^1)^{\epsilon_{T_1}(r_1+r_2)},
\label{eq:z2z2_wt0_GT3}\\
W_{\mathcal T}(r,\mu)
&=
(-1)^\mu \mathcal{W}_{\mathcal T},
\label{eq:z2z2_wt0_GT}\\
W_{S_4}(r,\mu)
&=
\eta_1^{\,f_s(r,\mu)}
(\tau^1)^{
\bar w_{S_4}\bar w_{\bar C_3}\epsilon_{T_1}f_s(r,\mu)
+
w_{\bar C_3}\epsilon_{S_4T_2}(r_2+r_3)
}
\nonumber\\
&\quad\times
(-1)^{
w_{S_4}\epsilon_{S_4T_2}
(\bar w_{\bar C_3}r_2+r_3)
}
\mathcal{W}_{S_4}(\mu)(i\tau^3)^{w_{S_4}},
\label{eq:z2z2_wt0_GS4}\\
W_{\bar C_3}(r,\mu)
&=
\eta_1^{\,f_c(r)}
(\tau^1)^{
\bar w_{S_4}\bar w_{\bar C_3}\epsilon_{T_1}f_c(r)
}
\nonumber\\
&\quad\times
(\tau^1)^{
\left(
w_{S_4}\bar w_{\bar C_3}\epsilon_{\bar C_3T_2}
+
w_{\bar C_3}\epsilon_{S_4T_2}
\right)(r_1+r_3)
}
\nonumber\\
&\quad\times
(-1)^{w_{\bar C_3}\epsilon_{S_4T_2}r_3}
\mathcal{W}_{\bar C_3}(\mu)(i\tau^3)^{w_{\bar C_3}} .
\label{eq:z2z2_wt0_GC3}
\end{align}
\end{subequations}

For $w_{\mathcal T}=1$, the final gauge representatives are
\begin{subequations}
\label{eq:z2z2_gauge_sol_wt1}
\begin{align}
W_{T_1}(r,\mu)
&=
\tau^0,
\label{eq:z2z2_wt1_GT1}\\
W_{T_2}(r,\mu)
&=
\eta_1^{\,r_1}
(\tau^1)^{\epsilon_{T_1}r_1},
\label{eq:z2z2_wt1_GT2}\\
W_{T_3}(r,\mu)
&=
\eta_1^{\,r_1+r_2}
(\tau^1)^{\epsilon_{T_1}(r_1+r_2)},
\label{eq:z2z2_wt1_GT3}\\
W_{\mathcal T}(r,\mu)
&=
\mathcal{W}_{\mathcal T}(\mu)i\tau^3,
\label{eq:z2z2_wt1_GT}\\
W_{S_4}(r,\mu)
&=
\eta_1^{\,f_s(r,\mu)}
(-1)^{
w_{S_4}\epsilon_{S_4T_2}
(\bar w_{\bar C_3}r_2+r_3)
}
\nonumber\\
&\quad\times
\mathcal{W}_{S_4}(\mu)(i\tau^3)^{w_{S_4}},
\label{eq:z2z2_wt1_GS4}\\
W_{\bar C_3}(r,\mu)
&=
\eta_1^{\,f_c(r)}
\mathcal{W}_{\bar C_3}(\mu)(i\tau^3)^{w_{\bar C_3}} .
\label{eq:z2z2_wt1_GC3}
\end{align}
\end{subequations}

The remaining gauge-inequivalent choices of the signs, the $\epsilon$ parameters, and the sublattice matrices $W_{\mathcal O}(\mu)$ are listed in Table~\ref{table:z2z2_psg}.  In total, we obtain $512$ algebraic PSGs of $\mathbb{Z}_2\times\mathbb{Z}_2$ type.

The labeling convention for the corresponding \textit{Ans\"atze} is
\begin{equation}
\label{eq:z2z2_label}
\mathrm{ZZA/B}\,w_{S_4\bar C_3}\,w_{\mathcal T},
\end{equation}
where the initial $A$ or $B$ denotes the translational class.
The label $w_{S_4\bar C_3}$ encodes the pair
$(w_{S_4},w_{\bar C_3})$ according to
\begin{equation}
\label{eq:z2z2_w_label}
w_{S_4\bar C_3}
=
\begin{cases}
0, & (w_{S_4},w_{\bar C_3})=(0,0),\\
1, & (w_{S_4},w_{\bar C_3})=(0,1),\\
2, & (w_{S_4},w_{\bar C_3})=(1,0),\\
3, & (w_{S_4},w_{\bar C_3})=(1,1).
\end{cases}
\end{equation}
The final digit is $w_{\mathcal T}=0$ or $1$.  The short-range
\textit{Ans\"atze} obtained from these PSGs, together with the nearby
$\mathbb{Z}_2$ descendants that appear upon including
third-neighbor amplitudes, are summarized in
Table~\ref{table:ansatz}.

\section{Zeroth-order mean-field Hamiltonian}
\label{app:zeroth_order}

In this appendix, we present the explicit zeroth-order mean-field Hamiltonian used in the self-consistency calculations.

For $S=1$ the Heisenberg exchange decouples exactly into the singlet channels, $J\,\mathbf S_i\cdot\mathbf S_j=-J:\!\big(\hat\chi^\dagger_{ij}\hat\chi_{ij}+\hat\eta^\dagger_{ij}\hat\eta_{ij}\big)\!:$~\cite{Liu-2010a}, which fixes $a(S=1)=1/2$ in Eq.~\eqref{eq:quadratic_ham}. The zeroth-order quadratic mean-field Hamiltonian is then
\begin{equation}
\label{eq:quadratic_ham_mean}
\begin{aligned}
\hat H^0_{\rm MF}
&=
\sum_{\langle ij\rangle}
J_{ij}\,
{\rm Tr}\!\left[
\tfrac12 u_{ij}^{\dagger}u_{ij}
-
\tfrac12\left(
 u_{ij}^{\dagger}\psi_i^{\dagger}\psi_j
+
{\rm H.c.}
\right)
\right]
\\
&\quad
+
\sum_i
\lambda_i\,
\big(\hat N_i-N_f\big)\,.
\end{aligned}
\end{equation}

For spin $1$, it is useful to work in a basis adapted to the $m=\pm1$ fermion flavors.  Following Ref.~\cite{Liu-2010a}, we introduce the symmetric and antisymmetric combinations
\begin{equation}
\label{eq:fs_fa_definition}
f_s
=
\frac{f_1+f_{-1}}{\sqrt{2}},
\qquad
f_a
=
\frac{f_1-f_{-1}}{\sqrt{2}} ,
\end{equation}
while keeping the $m=0$ fermion unchanged.  Equivalently,
\begin{equation}
\label{eq:fpm_inverse_definition}
f_1
=
\frac{f_s+f_a}{\sqrt{2}},
\qquad
f_{-1}
=
\frac{f_s-f_a}{\sqrt{2}} .
\end{equation}
This basis separates the symmetric and antisymmetric combinations of the $m=\pm1$ flavors and simplifies the diagonalization of Eq.~\eqref{eq:quadratic_ham_mean}.  The numerical self-consistency procedure is then carried out using the fermion flavors
\begin{equation}
f_s,\qquad f_a,\qquad f_0 .
\end{equation}
In this basis, Eq.~\eqref{eq:quadratic_ham_mean} becomes
\begin{widetext}
\begin{equation}\label{eq:decoupled Hamiltonian}
\hat H^0_{\rm MF}=E_0-
\sum_{\langle ij\rangle,\alpha}
\frac{J^{}_{ij}}{2}\,
\left[
\begin{pmatrix}
  f^\dagger_{i,\alpha} & p_\alpha f_{i,\alpha}
\end{pmatrix}
\begin{pmatrix}
\chi^{\star}_{ij} & -\eta_{ij}  \\
\eta^{\star}_{ij} & -\chi_{ij}
\end{pmatrix}
\begin{pmatrix}
 f_{j,\alpha} \\ p_\alpha f^\dagger_{j,\alpha}
\end{pmatrix}
+{\rm H.c.}
\right]
 +\sum_{i,\alpha}\frac{\lambda_i}{2}\begin{pmatrix}
  f^\dagger_{i,\alpha} & p_\alpha f_{i,\alpha}
\end{pmatrix}
\begin{pmatrix}
1 & 0  \\
0 & -1
\end{pmatrix}
\begin{pmatrix}
 f_{i,\alpha} \\ p_\alpha f^\dagger_{i,\alpha}
\end{pmatrix} ,
\end{equation}
\end{widetext}
where $\alpha\in\{s,a,0\}$ labels the three flavors, $p_s=-p_a=-p_0=1$, the matrix sandwiched between the Nambu spinors is the complex conjugate $u^{\ast}_{ij}$ of the link matrix in Eq.~\eqref{eq:link}, and
\begin{equation}
E_0=\sum_{\langle ij\rangle}J^{}_{ij}\left(|\chi_{ij}|^2+|\Delta_{ij}|^2\right)+\sum_i\lambda^{}_i\left(\tfrac{3}{2}-N_f\right).
\end{equation}
Therefore, the three flavors decouple into identical Bogoliubov--de~Gennes blocks, which is the origin of the overall factor of three in the self-consistency equations.

\section{Self-duality of the uniform-sector realizations of UA31 and UB31}
\label{app:selfdual}
\begin{figure*}[t]
    \includegraphics[width=1.0\linewidth]{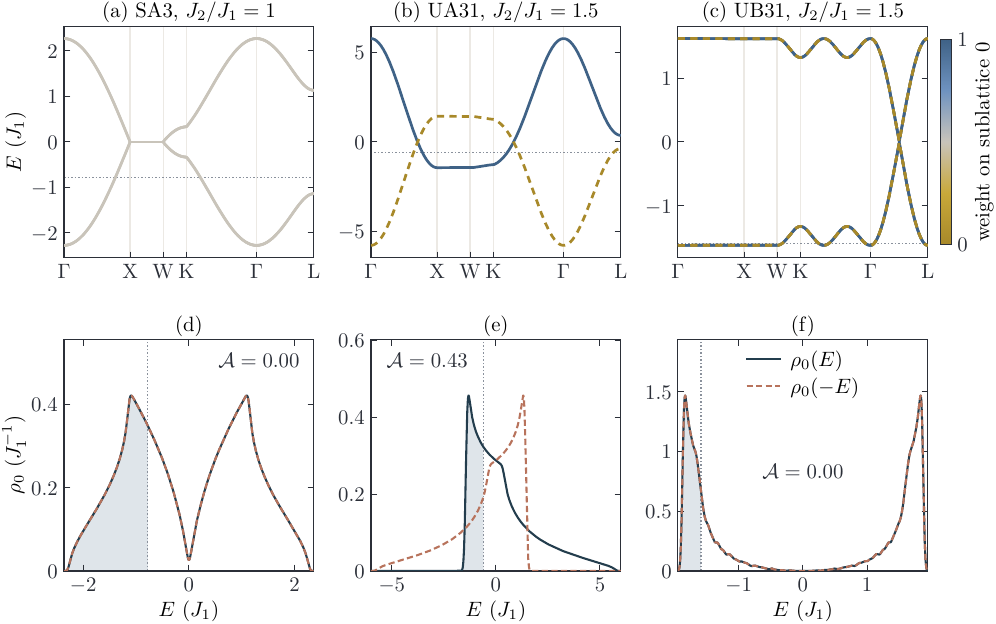}
    \caption{Self-duality of the $(1,1)$ realizations of the staggered-onsite \textit{Ans\"atze}, for the self-consistent parameters of Fig.~\ref{fig:energy_phase_diagram}(b). (a)--(c) Mean-field bands along $\Gamma$--X--W--K--$\Gamma$--L, colored by their weight on sublattice $0$; for the decoupled states (b) and (c) the bands of sublattice $0$ are solid and those of sublattice $1$ dashed. The dotted line is the Fermi level at filling $1/3$. (a) SA3, the $\chi^{}_2=\lambda=0$ realization of UA31 for $J_2/J_1\lesssim1.1$, at $J_2/J_1=1$; (b) UA31 with $\chi^{}_2\neq0$, $\chi^{}_1\simeq0$, at $J_2/J_1=1.5$; (c) the decoupled class-B realization of UB31 at $J_2/J_1=1.5$. (d)--(f) Density of states projected on sublattice $0$, $\rho_0(E)$, normalized to one, and its mirror image $\rho_0(-E)$; the occupied part is shaded. $\mathcal A=\frac12\int|\rho_0(E)-\rho_0(-E)|\,dE$ measures the departure from self-duality ($0\le\mathcal A\le1$). States (a) and (c) are self-dual and symmetric; state (b) is not and breaks the sublattice-exchanging operations $S_4$ and $\bar C_3$.}
    \label{fig:selfdual}
\end{figure*}

For UA31 and UB31 the second-neighbor amplitude and the onsite term are staggered, $u_{ij}=\pm \chi^{}_2\tau^3$ on the second-neighbor bonds of sublattices $0$ and $1$ and $\Lambda_i=\pm\lambda\tau^3$, and the gauge parts of the two sublattice-exchanging generators, $S_4$ and $\bar C_3$, lie in the $\tau^1$ coset of the IGG, whereas the translations act with gauge parts $\pm1$. Consider a realization at fixed sublattice fillings $(n_0,n_1)$. The lattice part of $S_4$ exchanges the sublattices and so maps the \textit{Ansatz} with $(\chi^{}_2,\lambda)$ at $(n_0,n_1)$ onto the \textit{Ansatz} with $(-\chi^{}_2,-\lambda)$ at $(n_1,n_0)$. Its gauge part $\tau^1$ reverses $(\chi^{}_2,\lambda)$ once more, restoring the \textit{Ansatz}, but it also maps $\zeta\to-\zeta$, i.e., $N_i\to3-N_i$. Altogether, $S_4$ maps the state with $(\chi^{}_2,\lambda)$ at $(n_0,n_1)$ onto the state with the same $(\chi^{}_2,\lambda)$ at $(3-n_1,3-n_0)$, and the same holds for $\bar C_3$. For $n_0+n_1=3$, i.e., $(3/2,3/2)$, $(1,2)$, and $(2,1)$, this is the state itself, which is why these realizations are symmetric (Table~\ref{table:ansatz-j1-j2}). In the uniform sector, $S_4$ maps $(1,1)$ onto $(2,2)$. Because the global $\tau^1$ is an exact duality between the two sectors, the $(2,2)$ state with $(\chi^{}_2,\lambda)$ is the same physical state as the $(1,1)$ state with $(-\chi^{}_2,-\lambda)$. A $(1,1)$ realization is therefore invariant under $S_4$ and $\bar C_3$ if and only if it is self-dual, i.e., if it coincides with the $(1,1)$ realization of the reversed amplitudes or, equivalently, if its particle and hole realizations coincide.

The criterion can be stated in terms of the single-particle spectrum. Let $H(\chi^{}_1,\chi^{}_2,\lambda)$ be the mean-field Hamiltonian of one flavor, and $W=\pm1$ on sublattice $0$ and $1$, a gauge transformation within $\mathrm{U}(1)$ that reverses $\chi^{}_1$. Then $H(\chi^{}_1,-\chi^{}_2,-\lambda)=-WH(\chi^{}_1,\chi^{}_2,\lambda)W$, so the dual $(1,1)$ state fills the highest third of the spectrum of $H$ where the original state fills the lowest third. Self-duality requires these two fillings to describe the same state. This is guaranteed if $H$ has a chiral symmetry, $\Gamma H\Gamma^{-1}=-H$, with $\Gamma$ a local gauge transformation, and is then visible in the density of states projected on one sublattice, $\rho_0(E)=\rho_0(-E)$. Figure~\ref{fig:selfdual} compares the bands and $\rho_0$ for the three cases that occur in Fig.~\ref{fig:energy_phase_diagram}(b); the decisive, gauge-invariant test is the comparison of the second-neighbor bond energies of the two sublattices, which $S_4$ exchanges.
\begin{itemize}
\item For $J_2/J_1\lesssim1.1$ the $(1,1)$ realizations of UA31 and UB31 collapse onto SA3 and SB1 ($\chi^{}_2=\lambda=0$). These are bipartite, $\Gamma=W$ is a chiral symmetry, and $\rho_0(E)=\rho_0(-E)$ [Fig.~\ref{fig:selfdual}(a),(d)]: the states are self-dual and symmetric.
\item For $J_2/J_1\ge1.2$ UA31 develops $\chi^{}_2\neq0$ and $\lambda\neq0$ with $\chi^{}_1\simeq0$, so that the two FCC sublattices decouple. Sublattice $0$ carries the second-neighbor band of the FCC lattice with hopping $+\chi^{}_2$, shifted by $+\lambda$, and sublattice $1$ its negative [Fig.~\ref{fig:selfdual}(b)]. The FCC lattice is not bipartite, its band is not symmetric about its center, and $\rho_0(E)\neq\rho_0(-E)$ [Fig.~\ref{fig:selfdual}(e)]. At filling $1/3$ the two sublattices realize physically different states: their second-neighbor bond energies are $-1.42$ and $-1.83$ (per unit cell, in units of $J_1$, at $J_2/J_1=1.5$), and the single amplitude $\chi_2=0.300$ is self-consistent only on average, the two sublattices separately requiring $0.263$ and $0.338$. The state is not self-dual and breaks $S_4$ and $\bar C_3$.
\item For $J_2/J_1\gtrsim1.2$ the $(1,1)$ realization of UB31 is the decoupled class-B state. Its sign structure makes the single-sublattice spectrum symmetric about $E=0$, the bands of the two sublattices coincide, and $\rho_0(E)=\rho_0(-E)$ [Fig.~\ref{fig:selfdual}(c),(f)]; the second-neighbor bond energies of the two sublattices agree within the mesh accuracy. The state is self-dual and symmetric.
\end{itemize}

Which symmetry is broken in the second case follows from the gauge parts. Translations and all products of an even number of $S_4$ and $\bar C_3$ have gauge parts in $\mathrm{U}(1)$ and preserve the constraint sector, whereas every element that exchanges the two FCC sublattices carries a gauge part in the $\tau^1$ coset. A non-self-dual $(1,1)$ realization therefore breaks exactly the sublattice-exchanging elements of the space group, reducing $Fd\bar 3m$ to its sublattice-preserving subgroup $F\bar43m$, the space group of the zincblende structure. This breaking of the sublattice-exchange symmetry should be distinguished from the chiral (``sublattice'') symmetry $\Gamma H\Gamma^{-1}=-H$ of the mean-field Hamiltonian. The latter is a property of the single-particle problem and not a symmetry of the spin model: it is lifted by $\chi^{}_2$ and $\lambda$ also for $N_f=(3/2,3/2)$ and $(1,2)$, where all states are symmetric. Its absence matters only in the uniform sectors, where it decides whether a staggered-onsite realization is self-dual.

\bibliography{arXiv_v1_spin1_PSG_Diamond}
\end{document}